\documentclass[hidelinks,a4paper,12pt,twoside]{book}

\usepackage{mathptmx} 
\usepackage{graphicx}
\usepackage{caption}

\usepackage{amsfonts}
\usepackage{amssymb}
\usepackage{amsmath}
\usepackage{multirow}
\usepackage{subfigure}
\usepackage{times}
\usepackage{url}

\usepackage{bookman}

\usepackage{fancyhdr}
\usepackage[figuresright]{rotating}
\usepackage{lscape}
\usepackage{calc}
\usepackage{float}                      
\usepackage{dcolumn}                    

\usepackage[breaklinks,colorlinks=false,pdfborder={0 0 0}, 
pdfauthor={Ashkbiz Danehkar}, pdftitle={Propagation of Electron-Acoustic Waves in a Plasma with Suprathermal Electrons}, pdfsubject={Plasma Physics}, pdfkeywords={Plasma Solitons, Plasma Waves, Electrostatic Waves, Dispersion Relations, Plasma Temperature}
]{hyperref}
\usepackage[hyphenbreaks]{breakurl}

\usepackage{color}
\usepackage{charter}

\renewcommand{\bibname}{References}

\definecolor{dark_grey}{gray}{0.3}
\definecolor{dark_grey2}{gray}{0.5}

\newcommand\getsto{\mathrel{\mathchoice {\vcenter{\offinterlineskip
\halign{\hfil
$\reset@font\displaystyle##$\hfil\cr\gets\cr\to\cr}}}
{\vcenter{\offinterlineskip\halign{\hfil$\reset@font\textstyle##$\hfil\cr\gets
\cr\to\cr}}}
{\vcenter{\offinterlineskip\halign{\hfil$\reset@font\scriptstyle##$\hfil\cr\gets
\cr\to\cr}}}
{\vcenter{\offinterlineskip\halign{\hfil$\reset@font\scriptscriptstyle##$\hfil\cr
\gets\cr\to\cr}}}}}

\newcommand\cor{\mathrel{\mathchoice {\hbox{$\widehat=$}}{\hbox{$\widehat=$}}
{\hbox{$\reset@font\scriptstyle\hat=$}}
{\hbox{$\reset@font\scriptscriptstyle\hat=$}}}}

\newcommand\lid{\mathrel{\mathchoice {\vcenter{\offinterlineskip\halign{\hfil
$\reset@font\displaystyle##$\hfil\cr<\cr\noalign{\vskip1.2pt}=\cr}}}
{\vcenter{\offinterlineskip\halign{\hfil$\reset@font\textstyle##$\hfil\cr<\cr
\noalign{\vskip1.2pt}=\cr}}}
{\vcenter{\offinterlineskip\halign{\hfil$\reset@font\scriptstyle##$\hfil\cr<\cr
\noalign{\vskip1pt}=\cr}}}
{\vcenter{\offinterlineskip\halign{\hfil$\reset@font\scriptscriptstyle##$\hfil\cr
<\cr
\noalign{\vskip0.9pt}=\cr}}}}}

\newcommand\gid{\mathrel{\mathchoice {\vcenter{\offinterlineskip\halign{\hfil
$\reset@font\displaystyle##$\hfil\cr>\cr\noalign{\vskip1.2pt}=\cr}}}
{\vcenter{\offinterlineskip\halign{\hfil$\reset@font\textstyle##$\hfil\cr>\cr
\noalign{\vskip1.2pt}=\cr}}}
{\vcenter{\offinterlineskip\halign{\hfil$\reset@font\scriptstyle##$\hfil\cr>\cr
\noalign{\vskip1pt}=\cr}}}
{\vcenter{\offinterlineskip\halign{\hfil$\reset@font\scriptscriptstyle##$\hfil\cr
>\cr
\noalign{\vskip0.9pt}=\cr}}}}}

\newcommand\sol{\mathrel{\mathchoice {\vcenter{\offinterlineskip\halign{\hfil
$\reset@font\displaystyle##$\hfil\cr\sim\cr<\cr}}}
{\vcenter{\offinterlineskip\halign{\hfil$\reset@font\textstyle##$\hfil\cr\sim\cr
<\cr}}}
{\vcenter{\offinterlineskip\halign{\hfil$\reset@font\scriptstyle##$\hfil\cr\sim\cr
<\cr}}}
{\vcenter{\offinterlineskip\halign{\hfil$\reset@font\scriptscriptstyle##$\hfil\cr
\sim\cr<\cr}}}}}

\newcommand\sog{\mathrel{\mathchoice {\vcenter{\offinterlineskip\halign{\hfil
$\reset@font\displaystyle##$\hfil\cr\sim\cr>\cr}}}
{\vcenter{\offinterlineskip\halign{\hfil$\reset@font\textstyle##$\hfil\cr\sim\cr
>\cr}}}
{\vcenter{\offinterlineskip\halign{\hfil$\reset@font\scriptstyle##$\hfil\cr
\sim\cr>\cr}}}
{\vcenter{\offinterlineskip\halign{\hfil$\reset@font\scriptscriptstyle##$\hfil\cr
\sim\cr>\cr}}}}}

\newcommand\lse{\mathrel{\mathchoice {\vcenter{\offinterlineskip\halign{\hfil
$\reset@font\displaystyle##$\hfil\cr<\cr\simeq\cr}}}
{\vcenter{\offinterlineskip\halign{\hfil$\reset@font\textstyle##$\hfil\cr
<\cr\simeq\cr}}}
{\vcenter{\offinterlineskip\halign{\hfil$\reset@font\scriptstyle##$\hfil\cr
<\cr\simeq\cr}}}
{\vcenter{\offinterlineskip\halign{\hfil$\reset@font\scriptscriptstyle##$\hfil\cr
<\cr\simeq\cr}}}}}

\newcommand\gse{\mathrel{\mathchoice {\vcenter{\offinterlineskip\halign{\hfil
$\reset@font\displaystyle##$\hfil\cr>\cr\simeq\cr}}}
{\vcenter{\offinterlineskip\halign{\hfil$\reset@font\textstyle##$\hfil\cr
>\cr\simeq\cr}}}
{\vcenter{\offinterlineskip\halign{\hfil$\reset@font\scriptstyle##$\hfil\cr
>\cr\simeq\cr}}}
{\vcenter{\offinterlineskip\halign{\hfil$\reset@font\scriptscriptstyle##$\hfil\cr
>\cr\simeq\cr}}}}}

\newcommand\grole{\mathrel{\mathchoice {\vcenter{\offinterlineskip\halign{\hfil
$\reset@font\displaystyle##$\hfil\cr>\cr\noalign{\vskip-1.5pt}<\cr}}}
{\vcenter{\offinterlineskip\halign{\hfil$\reset@font\textstyle##$\hfil\cr
>\cr\noalign{\vskip-1.5pt}<\cr}}}
{\vcenter{\offinterlineskip\halign{\hfil$\reset@font\scriptstyle##$\hfil\cr
>\cr\noalign{\vskip-1pt}<\cr}}}
{\vcenter{\offinterlineskip\halign{\hfil$\reset@font\scriptscriptstyle##$\hfil\cr
>\cr\noalign{\vskip-0.5pt}<\cr}}}}}

\newcommand\leogr{\mathrel{\mathchoice {\vcenter{\offinterlineskip\halign{\hfil
$\reset@font\displaystyle##$\hfil\cr<\cr\noalign{\vskip-1.5pt}>\cr}}}
{\vcenter{\offinterlineskip\halign{\hfil$\reset@font\textstyle##$\hfil\cr
<\cr\noalign{\vskip-1.5pt}>\cr}}}
{\vcenter{\offinterlineskip\halign{\hfil$\reset@font\scriptstyle##$\hfil\cr
<\cr\noalign{\vskip-1pt}>\cr}}}
{\vcenter{\offinterlineskip\halign{\hfil$\reset@font\scriptscriptstyle##$\hfil\cr
<\cr\noalign{\vskip-0.5pt}>\cr}}}}}

\newcommand\loa{\mathrel{\mathchoice {\vcenter{\offinterlineskip\halign{\hfil
$\reset@font\displaystyle##$\hfil\cr<\cr\approx\cr}}}
{\vcenter{\offinterlineskip\halign{\hfil$\reset@font\textstyle##$\hfil\cr
<\cr\approx\cr}}}
{\vcenter{\offinterlineskip\halign{\hfil$\reset@font\scriptstyle##$\hfil\cr
<\cr\approx\cr}}}
{\vcenter{\offinterlineskip\halign{\hfil$\reset@font\scriptscriptstyle##$\hfil\cr
<\cr\approx\cr}}}}}

\newcommand\goa{\mathrel{\mathchoice {\vcenter{\offinterlineskip\halign{\hfil
$\reset@font\displaystyle##$\hfil\cr>\cr\approx\cr}}}
{\vcenter{\offinterlineskip\halign{\hfil$\reset@font\textstyle##$\hfil\cr
>\cr\approx\cr}}}
{\vcenter{\offinterlineskip\halign{\hfil$\reset@font\scriptstyle##$\hfil\cr
>\cr\approx\cr}}}
{\vcenter{\offinterlineskip\halign{\hfil$\reset@font\scriptscriptstyle##$\hfil\cr
>\cr\approx\cr}}}}}

\newcommand\diameter{{\ifmmode\mathchoice
{\ooalign{\hfil\hbox{$\reset@font\displaystyle/$}\hfil\crcr
{\hbox{$\reset@font\displaystyle\mathchar"20D$}}}}
{\ooalign{\hfil\hbox{$\reset@font\textstyle/$}\hfil\crcr
{\hbox{$\reset@font\textstyle\mathchar"20D$}}}}
{\ooalign{\hfil\hbox{$\reset@font\scriptstyle/$}\hfil\crcr
{\hbox{$\reset@font\scriptstyle\mathchar"20D$}}}}
{\ooalign{\hfil\hbox{$\reset@font\scriptscriptstyle/$}\hfil\crcr
{\hbox{$\reset@font\scriptscriptstyle\mathchar"20D$}}}}
\else{\ooalign{\hfil/\hfil\crcr\mathhexbox20D}}%
\fi}}

\newcommand\sq{\ifmmode\squareforqed\else{\unskip\nobreak\hfil
\penalty50\hskip1em\null\nobreak\hfil\squareforqed
\parfillskip=0pt\finalhyphendemerits=0\endgraf}\fi}
\newcommand\squareforqed{\hbox{\rlap{$\sqcap$}$\sqcup$}}

\newcommand{\romn}[1] {{\mathrm #1}}

\newcommand\fp{\hbox{$.\!\!^{\reset@font\reset@font\scriptscriptstyle\romn p}$}}

\renewcommand{\chaptername}[1]{\fontfamily{ptm}\selectfont}

\makeatletter
\def\@endpart{\vfil\newpage}
\makeatother

\newenvironment{abs}[2]{
  \newpage
  \thispagestyle{plain}
  \markboth{}{}
  \begin{center}
 {\huge\bf #2}
  \end{center}
  \vspace{10mm}}{
  \clearpage}

\def\p0{\phantom{0}}

\def\lessim{\raise-.5ex\hbox{$\buildrel<\over{\scriptstyle\mathtt{\sim}}$}}
\def\grtsim{\raise-.5ex\hbox{$\buildrel>\over{\scriptstyle\mathtt{\sim}}$}}

\begin{document}


\frontmatter
\begin{titlepage}
\pagenumbering{roman}
\clearpage
\thispagestyle{empty}
\vspace*{10mm}

\begin{center}


\vspace*{20mm}

{\Huge Propagation of \linebreak  Electron-Acoustic Waves  \linebreak 
in a Plasma with \linebreak  Suprathermal Electrons
}

\vspace*{30mm}

{\Large \bf
Ashkbiz Danehkar}

\vspace*{10mm}

A thesis submitted to \linebreak
the Queen's University Belfast
\linebreak for the Degree of Master of Science \linebreak
in Plasma Physics

\vspace*{20mm}



Centre for Plasma Physics\linebreak
Department of Physics \& Astronomy\linebreak
Queen's University Belfast\linebreak
Belfast BT7 1NN, United Kingdom
\linebreak\linebreak
December 2009

\end{center}
\clearpage
\thispagestyle{empty}

\end{titlepage}



\newpage
\begin{center}


\end{center}

\cleardoublepage


\begin{abs}{Abstract}{\Large Abstract}

\noindent


Electron-acoustic waves occur in space and laboratory plasmas where two distinct electron populations exist, namely cool and hot electrons. The observations revealed that the hot electron distribution often has a long-tailed suprathermal (non-Maxwellian) form. The aim of the present study is to investigate how various plasma parameters modify the electron-acoustic structures. We have studied the electron-acoustic waves in a collisionless and unmagnetized plasma consisting of cool inertial electrons, hot suprathermal electrons, and mobile ions. First, we started with a cold one-fluid model, and we extended it to a warm model, including the electron thermal pressure. Finally, the ion inertia was included in a two-fluid model. The linear dispersion relations for electron-acoustic waves depicted a strong dependence of the charge screening mechanism on excess suprathermality. A nonlinear (Sagdeev) pseudopotential technique was employed to investigate the existence of electron-acoustic solitary waves, and to determine how their characteristics depend on various plasma parameters. The results indicate that the thermal pressure deeply affects the electron-acoustic solitary waves. Only negative polarity waves were found to exist in the one-fluid model, which become narrower as deviation from the Maxwellian increases, while the wave amplitude at fixed soliton speed increases. However, for a constant value of the true Mach number, the amplitude decreases for increasing suprathermality. It is also found that the ion inertia has a trivial role in the supersonic domain, but it is important to support positive polarity waves in the subsonic domain.

\end{abs}

\newpage
\mbox{}
\newpage

\addcontentsline{toc}{section}{Abstract}
\cleardoublepage


\begin{abs}{Declaration}{\Large Declaration}

\noindent I confirm the following:\newline

\noindent (i) the dissertation is not one for which a degree has been or will be conferred by any other university
or institution;\newline

\noindent (ii) the dissertation is not one for which a degree has already been conferred by this university;\newline

\noindent (iii) that this work submitted for assessment is my own and expressed in my own words. Any use
made within it of works of other authors in any form (e.g. ideas, figures, text, tables) are properly
acknowledged at their point of use. A list of the references employed is included;\newline

\noindent (iv) the composition of the dissertation is my own work.

    \vspace{1.5cm}


\noindent Ashkbiz Danehkar\\
\noindent December 2009

\end{abs}



\begin{abs}{Acknowledgments}{\Large Acknowledgments}

\noindent

The author would like to thank Dr. Ioannis Kourakis, Lecturer at the Queen's University Belfast for his supervision and guidance throughout the project, Dr. Nareshpal Singh Saini, Post-doctoral Researcher at the Queen's University Belfast, Prof. Manfred A. Hellberg,  Emeritus Professor  at the University of KwaZulu-Natal, for their collaborations, providing consultations and valuable comments. This research has been
supported by a grant from the Department for Employment and Learning (DEL) in
Northern Ireland (A.D.), and a contract from the Engineering and Physical
Sciences Research Council (UK EPSRC) (N.S.S. \& I.K.). The research is also
supported in part by the National Research Foundation of South Africa (NRF)
(M.A.H.). Any opinion, findings, and conclusions or recommendations expressed
in this material are those of the authors and therefore the NRF does not
accept any liability in regard thereto.

\end{abs}
\cleardoublepage

\tableofcontents
\addcontentsline{toc}{section}{Contents}
\cleardoublepage

\listoffigures
\addcontentsline{toc}{section}{List of Figures}
\newpage








\mainmatter
\clearpage
\newpage
\pagenumbering{arabic}
%
%
%


\chapter{Introduction}\label{ch:intro}

The electron-acoustic waves (EAWs) usually occur in a plasma, where inertial
cool electrons oscillates against inertialess hot electrons. EAWs may exist in
plasmas with two electrons population referred to as cool\,\footnote{We distinguished \textquotedblleft cool\textquotedblright ($T_{c}/T_{h}\ll1$) from \textquotedblleft cold\textquotedblright ($T_{c}/T_{h}=0$).}
(hot) electrons with respective temperatures $T_{c}$ ($T_{h}$). These are
typically high-frequency (in comparison with the ion plasma frequency)
electrostatic waves propagating with the phase velocity intermediate between
hot and cool electron thermal velocities. At such high frequency, the positive
ions behave like uniformly distributed charge background providing charge
neutrality, but they have no essential role in the dynamics (of supersonic
negative solitary waves; see
\ref{twofluidmodel:nonlinear:rarefactiveexistence}). The phase velocity of the
EAWs is much larger than the cool electron thermal velocity and much smaller
than the hot electron thermal velocity. The cool electrons provide the
inertial effects needed to maintain the EAWs, while the restoring force comes
from the pressure of the hot electrons.

\begin{figure}
\begin{center}
\includegraphics[width=4in]{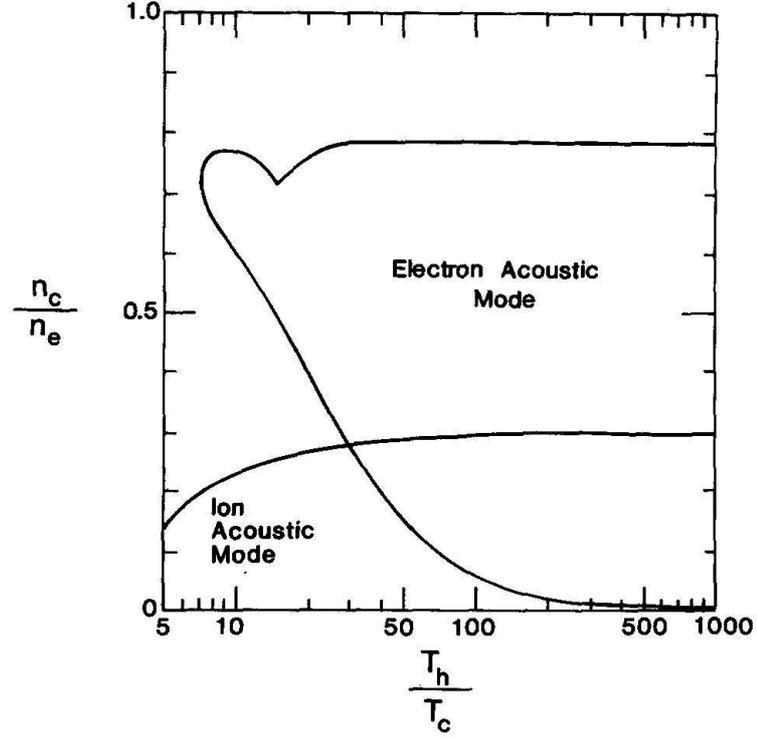}
\caption[The parameter space of the cool-to-total electron density ratio
vs. the hot-to-cool temperature ratio for weakly damped EAWs and IAWs]{The parameter space of the cool-to-total electron density ratio
versus the hot-to-cool temperature ratio for weakly damped electron-acoustic
waves and ion-acoustic waves \cite{Gary1985}.}
\label{fig1_19}
\end{center}
\end{figure}

As the temperature rises in a collisionless plasma, the phase velocity of
waves become comparable with the electron thermal velocities. In a situation
depends on the electron thermal velocity (faster/slower than the phase
velocity), a direct interaction between electrons and waves produces the
Landau damping (wave heating) or inverse Landau damping (instabilities)
through the Vlasov kinetic theory (no need of a collision term). When the
phase velocity goes near the thermal velocity for a short wavelength, the
Landau damping become very strong, i.e., the wave cannot propagate in the
plasma. This means that the propagation of EAWs is possible within a
restricted range of parameters. It has been proven that the EAWs are not
damped at the temperature ratio $T_{c}/T_{h}\lesssim0.1$
\cite{Berthomier1999,Gary1985} and the cool electrons at a significant
fraction of the total electron density: $0.2\lesssim n_{c}/(n_{c}+n_{h})\lesssim0.8$ \cite{Gary1985,Tokar1984}, as illustrated in Fig.
\ref{fig1_19}. The wave number $k$ of the weakly damped EAW is between roughly
$0.2\lambda_{Dc}^{-1}$ and $0.6\lambda_{Dc}^{-1}$ (where $\lambda_{Dc}$ is the
cool electron Debye length). The temperature and the number density of the
cool and hot electrons modify the stable range of the wave number (see e.g.
Figs. 1--3 in \cite{Gary1985}; or Fig. \ref{fig1_1}(b) and Fig. \ref{fig2_2}).

The EAWs often occur in laboratory experiments
\cite{Derfler1969,Henry1972,Ikezawa1981} and space plasmas e.g. the Earth's
bow shock \cite{Thomsen1983,Feldman1983,Bale1998} and the auroral
magnetosphere \cite{Tokar1984,Lin1984}. Another example is the Broadband
Electrostatic Noise (BEN), a common wave activity in the plasma sheet boundary
layer (PSBL) region, which has been observed by the satellites
missions \cite{Matsumoto1994,Franz1998,Cattell1999,Kakad2009}. The BEN
emissions forming as EAWs, which include a series of isolated bipolar pulses,
have the frequency range from $\sim10$ Hz upto the local electron plasma
frequency ($\sim10$ kHz) \cite{Matsumoto1994}. This suggests that the
emissions are related to electron dynamics rather than ions
\cite{Matsumoto1994,Kakad2009}.

In two electron temperature plasmas, two electrons population are often
characterized by a thermal Maxwellian distribution
\cite{Nishihara1981,Leubner1982,Armstrong1983,Berthomier1998}. However, some
space and laboratory plasmas have such a suprathermal electron population,
whose behaviors are extremely different from a Maxwellian distribution.
Electrons obey an inverse power law distribution at a velocity much higher
than the electron thermal velocity. We describe this suprathermal population
by a generalized Lorentzian or $\kappa$-distributions
\cite{Schippers2008,Hellberg2000,Baluku2008}.

The common form of  the isotropic (three-dimensional) generalized Lorentzian
or $\kappa$-distribution function is given by
\cite{Hellberg2000,Baluku2008,Saini2009}
\begin{equation}
f_{\kappa}(v)=n_{0}(\pi\kappa\theta^{2})^{-3/2}\frac{\Gamma(\kappa+1)}
{\Gamma(\kappa-\frac{1}{2})}\left(  1+\frac{v^{2}}{\kappa\theta^{2}}\right)
^{-\kappa-1}. \label{eq1_60}
\end{equation}
where $n_{0}$ is an equilibrium number density of the electrons, $v$ the
species velocity, $\theta$ is a generalized thermal velocity related to the
actual thermal velocity of the electrons $v_{th,e}=(2k_{B}T_{e}/m_{e})^{1/2}$
by $\theta=v_{th,e}\left(  (\kappa-\tfrac{3}{2})/\kappa\right)  ^{1/2}$;
$k_{B}$ the Boltzmann constant, $m_{e}$ and $T_{e}$ the mass and temperature
of the electrons, respectively. We note that $\kappa$ is the spectral index
of $\kappa$-distributions with $\kappa>\frac{3}{2}$. For $\kappa
\rightarrow\infty$, we have a Maxwellian, while low values of $\kappa$ are
associated with significant numbers of suprathermal particles. The gamma
function $\Gamma$ arises from the normalization of $f_{\kappa}(v)$, i.e. $\int
f_{\kappa}(v)d^{3}v=n_{0}$.

The $\kappa$-distribution has been firstly applied to model velocity
distribution of particles observed in space plasmas, often in the range
$2<\kappa<6$ \cite{Vasyliunas1968}. The $\kappa$-distribution function can
describe laboratory experiments and space plasmas more effectively than a
Maxwellian function
\cite{Feldman1983,Schippers2008,Christon1988,Mace1995,Abbasi2007,Pierrard1996}. 
For example, measurements of plasma sheet electron and ion distributions can
be treated by $\kappa_{i}=4.7$ and $\kappa_{e}=$ $5.5$ \cite{Christon1988}
(here, $e$ denotes electrons and $i$ ions), observations in the earth's
foreshock satisfying $3<\kappa_{e}<6$ \cite{Feldman1983}, and coronal
electrons in solar wind model with $2<\kappa_{e}<6$ \cite{Pierrard1996}.

Studies of linear and nonlinear EAWs in plasmas with nonthermal electrons have
received a great deal of interest in recent years
\cite{Mace1995,Summers1991,Mace1999,Hellberg2002,Dubouloz1991,Berthomier2000,Singh2004,Kourakis2004,Gill2006,Saini2009}. 
The linear analysis of EAWs, which provided a dispersion relation, was
firstly described in an unmagnetized homogenous plasma \cite{Fried1961}. It
exhibited a heavily damped acoustic-like solution in addition to the Langmuir
waves and ion-acoustic waves (IAWs) \cite{Fried1961}. The linear properties
with suprathermal particles provided dispersion functions
\cite{Mace1995,Summers1991,Mace1999,Hellberg2002}.  It shows the effect of
suprathermal electrons on propagation of EAWs, which increase the Landau
damping of the wave at small wave numbers (acoustic regime) \cite{Mace1999},
and the dependence of the Landau damping on the fraction of suprathermal
electrons \cite{Hellberg2002}. Large values of $\kappa$ (quasi-Maxwellian)
produce weaker Landau damping in the acoustic regime, while Landau damping
increases by hot electrons for small values of $\kappa$ \cite{Mace1999}.

The nonlinear analysis of the EAWs in a one-dimensional unmagnetized plasma
composed of cold and hot electrons has been shown the existence of negative
potential soliton \cite{Dubouloz1991}, while additional electron beam
component leads to a positive potential soliton \cite{Berthomier2000}. The
nonlinear aspects of EAWs in an unmagnetized plasma consisting of nonthermal
electrons, fluid cold electrons, and ions provided negative potential solitary
structures \cite{Singh2004}.

\section{Thesis Outline}\label{ch:intro:outline}

We used a strategic workplan and some steps for this work. The analytical
basis for the 3 models is presented in the Appendix \ref{workplan}. We discuss
the outcomes of each model for the linear dispersion relation and the
existence conditions of stationary profile solitary structures. The
organization of the thesis is as follows:

In Chapter \ref{singlecold}, we have performed a preliminary work on a
one-fluid cold ($T_{c}=0$) model consisting of cold electron and background of
hot suprathermal electrons and stationary ions, i.e., only cold electrons
treated as a fluid. We study the linear and nonlinear effects of the hot
suprathermal electrons on electron-acoustic (EA) waves, namely the weakly
damped region and the propagation velocity range.

In Chapter \ref{warmmodel}, we extended it to the one-fluid warm
electrons model, which includes the pressure of the cool ($T_{c}\neq0$)
electrons. Comparing with the one-fluid cold model, we investigate the effect
of the \textquotedblleft cool-electron\textquotedblright temperature. We
distinguish two regimes for the propagation velocity, namely subsonic (slow)
and supersonic (fast) scales. We have treated the cool electrons to be
supersonic (i.e. having a propagation speed above the electron thermal speed),
and have found that only negative solitary structures can exist on this (fast) scale.

In Chapter \ref{twofluidmodel}, we assume that ions are no longer
stationary, i.e., treated as a fluid to make a two-fluid model consisting of
cool electron-fluid, ion-fluid, and hot background of suprathermal electrons.
We see that the ion-fluid does not influence much the fast negative solitons,
while producing novel positive solitary structures on the slow scale. We also
investigate the nonlinear effects of the hot suprathermal electrons on the
positive acoustic solitary waves, i.e., the electric potential pulse and the
propagation velocity range.

Finally, our main findings and conclusions are summarized in Chapter \ref{conclusion}. 








%
%
%


\chapter{Cold Electron Fluid with Suprathermal Electrons}\label{singlecold}

In this chapter, we study the EAWs in an unmagnetized plasma composed of cold
($T_{c}=0$) electron fluid, hot suprathermal $\kappa$-distributed electrons,
and uniformly distributed ions. We present the basic set of equations of the
model in \S  \ref{singlecold:basic}. In \S  \ref{singlecold:dr}, we derive
the dispersion relation for the linear dynamics of EAWs. In
\S  \ref{singlecold:nonlinear}, we obtain the nonlinear structures of the
electrostatic solitary waves and describe the soliton existence domain.

\section{Basic Equations}\label{singlecold:basic}

We consider a plasma with three components, namely cold electron-fluid,
inertialess hot electron component with a suprathermal (non-Maxwellian)
electron velocity distribution, and uniform ion background. The cold
electron-fluid governing the linear and nonlinear dynamics of
electron-acoustic waves (EAWs) feels the effect of the hot suprathermal
electrons. To study the linear and nonlinear results, we obtain the normalized
fluid-moment equations and the Poisson's equation through some appropriate scales.

The number density of cold electrons is governed by the continuity equation
\begin{equation}
\frac{\partial n_{c}}{\partial t}+\frac{\partial(n_{c}u_{c})}{\partial x}=0.
\label{eq1_2}
\end{equation}
The cold electrons obey the momentum equation
\begin{equation}
\frac{\partial u_{c}}{\partial t}+u_{c}\frac{\partial u_{c}}{\partial x}
=\frac{e}{m_{e}}\frac{\partial\phi}{\partial x}. \label{eq1_3}
\end{equation}
The densities of suprathermal hot electron, fluid cold electrons and uniform
ions are related to the electrostatic potential by the Poisson's equation:
\begin{equation}
\frac{\partial^{2}\phi}{\partial x^{2}}=-\frac{e}{\varepsilon_{0}}\left(
n_{i}-n_{c}-n_{h}\right)  , \label{eq1_4}
\end{equation}
where $\varepsilon_{0}$ is the permittivity constant.

The uniform ions mean that $n_{i}=n_{i,0}=$ const, where $n_{i0}$ is the
undisturbed ion density. We need an expression for the number density of the
hot electron, $n_{h}$, which takes into account the suprathermal distribution
(\ref{eq1_60}). Integrating Eq. (\ref{eq1_60}) over velocity space, we obtain
the number density of the suprathermal hot electrons given by
\cite{Baluku2008}
\begin{equation}
n_{h}(\phi)=n_{h,0}\left(  1-\frac{e\phi}{k_{B}T_{h}(\kappa-\tfrac{3}{2}
)}\right)  ^{-\kappa+1/2} , \label{eq1_1}
\end{equation}
where $n_{h,0}$ is the density of hot electrons in the undisturbed plasma,
$T_{h}$ the temperature of hot electron, $\phi$ the electrostatic wave
potential, $e$ the elementary charge, and $\kappa$ a spectral index which
measures deviation from thermal equilibrium.

At equilibrium, the plasma is assumed to be quasi-neutral
\begin{equation}
n_{c,0}+n_{h,0}=Zn_{i,0}. \label{eq1_5}
\end{equation}
In addition, we define the equilibrium density ratios of the ions to the cold
electrons and of the hot electrons to the cold electrons, respectively:
\begin{equation}
\begin{array}
[c]{cc}
\alpha\equiv\dfrac{n_{i,0}}{n_{c,0}}, & \text{  }\beta\equiv\dfrac{n_{h,0}
}{n_{c,0}}.
\end{array}
\label{eq1_8}
\end{equation}
We assume that $Z=1$ everywhere. Using above definition, Eq. (\ref{eq1_5})
take the form as $\alpha=1+\beta$. According to \cite{Tokar1984}, the
propagation of the EAWs remain undamped in the range $0.2\lesssim
n_{c,0}/(n_{c,0}+n_{h,0})\lesssim0.8$. Therefore, the following condition is
satisfied: $0.25<\beta<4$. This is a range for the existence of
electron-acoustic solitary waves.

If we scale densities by $n_{c,0}$, we can write Eq. (\ref{eq1_1}) in
dimensionless form as
\begin{equation}
n_{h}(\phi)=\beta\left(  1-\frac{\phi}{\kappa-\tfrac{3}{2}}\right)
^{-\kappa+1/2}. \label{eq1_63}
\end{equation}
In the limit $\kappa\rightarrow\infty$, Eq. (\ref{eq1_63}) is reduced to
$n(\phi)=\beta\exp(\phi)$, the Maxwellian distributions for the electrons.

It is convenient to use the nondimensional form of Eqs. (\ref{eq1_2})--(\ref{eq1_4}):
\begin{equation}
\frac{\partial n}{\partial t}+\frac{\partial(nu)}{\partial x}=0,
\label{eq1_10}
\end{equation}
\begin{equation}
\frac{\partial u}{\partial t}+u\frac{\partial u}{\partial x}=\frac
{\partial\phi}{\partial x}, \label{eq1_11}
\end{equation}
\begin{equation}
\frac{\partial^{2}\phi}{\partial x^{2}}=-(1+\beta)+n+\beta\left(  1-\frac
{\phi}{\kappa-\tfrac{3}{2}}\right)  ^{-\kappa+1/2}, \label{eq1_12}
\end{equation}
which is done by choosing the variables as
\begin{equation}
\begin{array}
[c]{ccccc}
\dfrac{n_{c}}{n_{c,0}}\rightarrow n,\text{ } & \dfrac{\phi}{k_{B}T_{h}
/e}\rightarrow\phi,\text{ } & \dfrac{u_{c}}{c_{h,s}}\rightarrow u,\text{ } &
t\omega_{pc}\rightarrow t,\text{ } & \dfrac{x}{\lambda_{0}}\rightarrow x,
\end{array}
\label{eq1_7}
\end{equation}
where the sound speed of hot electrons is defined by $c_{h,s}=(k_{B}T_{h}/m_{e})^{1/2}$, the plasma frequency of cold electrons $\omega
_{pc}=(n_{c,0}e^{2}/\varepsilon_{0}m_{e})^{1/2}$, and a characteristic length
scale $\lambda_{0}=(\varepsilon_{0}k_{B}T_{h}/n_{c,0}e^{2})^{1/2}$.

\section{Linear Dispersion Relation}\label{singlecold:dr}

In this section, we use linear analysis to derive the dispersion relation for
the linear dynamics of EAWs. The linear dispersion relation exhibits that the
frequency of the EAWs are less than the cold electron plasma frequency and in
the long-wavelength mode the EAWs behave like an ion-acoustic wave.

Let $S$ be any of the system variables $n$, $u$, and $\phi$, describing the
system's state at a given position $x$ and instant $t$. We shall consider
small deviations from the equilibrium state $S^{(0)}$, explicitly $n^{(0)}=1$,
$u^{(0)}=0$ and $\phi^{(0)}=0$, by taking
\begin{equation}
S=S^{(0)}+S_{1}^{(1)}e^{i(kx-\omega t)}. \label{eq1_15}
\end{equation}
Accordingly, the derivatives of the first order amplitudes are treated as
\begin{equation}
\begin{array}
[c]{cc}
\dfrac{\partial S_{1}^{(1)}}{\partial t}=-i\omega S_{1}^{(1)},\text{  } &
\dfrac{\partial S_{1}^{(1)}}{\partial x}=ikS_{1}^{(1)}.
\end{array}
\label{eq1_17}
\end{equation}

Eqs. (\ref{eq1_10}) and (\ref{eq1_11}) lead to the following expressions for
density and velocity in terms of potential, namely
\begin{equation}
\begin{array}
[c]{cc}
n_{1}^{(1)}=-\dfrac{k^{2}}{\omega^{2}}\phi_{1}^{(1)},\text{  } & u_{1}
^{(1)}=-\dfrac{k}{\omega}\phi_{1}^{(1)}.
\end{array}
\label{eq1_18}
\end{equation}
where $\omega$ is the wave frequency and $k$ the wavenumber.

Similarly, the Poisson's equation (\ref{eq1_12}) provides the compatibility
condition as
\begin{equation}
-k^{2}\phi_{1}^{(1)}=-\beta-\frac{k^{2}}{\omega^{2}}\phi_{1}^{(1)}
+\beta\left(  1-\frac{\phi_{1}^{(1)}}{\kappa-\tfrac{3}{2}}\right)
^{-\kappa+1/2}. \label{eq1_19}
\end{equation}
Let us make use of the expansion keeping up to first order:
\begin{equation}
n_{h}(\phi)\approx1+\frac{\kappa-\frac{1}{2}}{\kappa-\tfrac{3}{2}}\phi.
\label{eq1_20}
\end{equation}
Using above approximate relation, Eq. (\ref{eq1_19}) provide the familiar EAWs
dispersion relation:
\begin{equation}
\omega_{1}^{2}=\frac{k^{2}}{k^{2}+k_{D}^{2}}, \label{eq1_21}
\end{equation}
where we define $k_{D}$ as
\begin{equation}
k_{D}\equiv\dfrac{1}{\lambda_{D}}\equiv\left(  \dfrac{\beta(\kappa-\frac{1}
{2})}{\kappa-\tfrac{3}{2}}\right)  ^{1/2}. \label{eq1_22}
\end{equation}

Restoring dimensions, we get the standard dispersion relation
\begin{equation}
\omega_{1}^{2}=\omega_{pc}^{2}\frac{k^{2}\lambda_{Dh}^{2}}{k^{2}\lambda
_{Dh}^{2}+\left(  \dfrac{\kappa-\frac{1}{2}}{\kappa-\tfrac{3}{2}}\right)  },
\label{eq1_23}
\end{equation}
where $\lambda_{Dh}$ is the (hot electron) Debye length defined by
\begin{equation}
\lambda_{Dh}=\left(  \dfrac{\varepsilon_{0}k_{B}T_{h}}{n_{h,0}e^{2}}\right)
^{1/2}=\beta^{-1/2}\lambda_{0}. \label{eq1_23_1}
\end{equation}
Eq. (\ref{eq1_23}) is recognized as the linear dispersion equation governing
our model. This can be represented as curves on a $k$--$\omega$ plane, as
dimensionless dispersion relation (\ref{eq1_21}) shown in Fig. \ref{fig1_1}.
It is important that the EAWs will be deeply damped for the wave number $k$
greater than $0.6k_{D}$. Particularly, the linear EAWs are weakly damped
between roughly $0.2k_{D}$ and $0.6k_{D}$ \cite{Gary1985,Tokar1984}. The
stable range of the wave number rises with growing the equilibrium density
ratio $\beta=n_{h,0}/n_{c,0}$. The linear EAWs (unlike the well-known Langmuir
waves) extends only up to the cold electron plasma frequency. On the other
hand, the dispersion relation in the long-wavelength limit (in comparison with
$\lambda_{Dh}$) is $\omega\simeq kC_{s}$ where $C_{s}$ is the
electron-acoustic sound speed given by
\begin{equation}
C_{s}=\beta^{-1/2}\left(  \frac{\kappa-\tfrac{3}{2}}{\kappa-\frac{1}{2}
}\right)  ^{1/2}c_{h,s}. \label{eq1_23_2}
\end{equation}
The long-wavelength mode is analogous to an ion-acoustic (IA) mode. Here, the
cold electron plays the role of cold ions in the IA mode.

\subsection{Hot Suprathermal Effect on Linear Waves}

As the temperature of the hot electrons is increased, the sound speed within
the range of the long-wavelength increases. But, increasing $\beta $or
decreasing $\kappa$ reduces the sound speed.

\begin{figure}
\begin{center}
\includegraphics[width=6.0in]{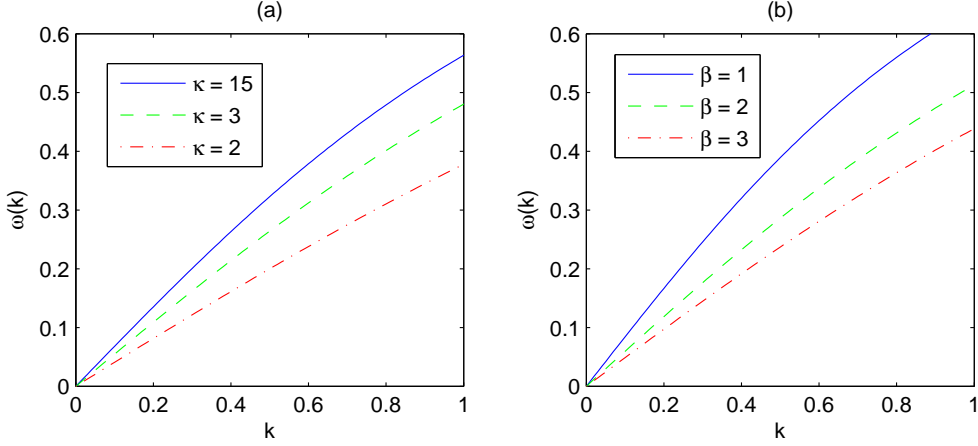}
\caption[Dispersion curves for the linear EAWs.]{Dispersion curves for the linear EAWs. (a) Variation of the
dispersion function curve for different values of  $\kappa$ and $\beta=2$.
Curves from bottom to top: $\kappa=15$ (solid), $3$ (dashed), $2$ (dot-dashed curve). (b) Variation of the dispersion function curve for
different values of  $\beta$ and $\kappa=4$. Curves from top to bottom:
$\beta=1$ (solid), $2$ (dashed), $3$ (dot-dashed curve).}
\label{fig1_1}
\end{center}
\end{figure}

As illustrated in Fig. \ref{fig1_1}, the dispersion curve depends on the
parameters $\kappa$ and $\beta$. In the weakly damped region ($0.25<\beta<4$),
the slope of the dispersion curve rises with either the increase in $\kappa$
or the decrease in $\beta$. Thus, growing $\beta$ broadens the range of
permitted frequencies, within the weakly damped region ($0.2k_{D}<k<0.6k_{D}$).

\section{Nonlinear Electron-Acoustic Solitary Waves}\label{singlecold:nonlinear}

Now, we employ the Sagdeev pseudopotential approach \cite{Sagdeev1966} to
investigate the nonlinear propagation properties of the cold electrons in a
plasma under the effect of the hot suprathermal electrons. In
\S \ref{singlecold:nonlinear:existence}, we discuss necessary conditions for
the generation of solitary structures in the plasma.

We consider solutions of Eqs. (\ref{eq1_10})--(\ref{eq1_12}), that are
stationary in a frame moving with velocity $M$. We use the Galilean
transformation, $\xi=x-Mt$ and $\tau=t$, where $M$ is called the Mach number.
This means that all derivatives shall be replaced as follows
\begin{equation}
\begin{array}
[c]{cc}
\dfrac{\partial}{\partial x}=\dfrac{d}{d\xi},\text{  } & \dfrac{\partial
}{\partial t}=-M\dfrac{d}{d\xi}.
\end{array}
\label{eq1_40}
\end{equation}
Therefore, Eqs. (\ref{eq1_10})--(\ref{eq1_12}) take the following form
\begin{equation}
-M\dfrac{dn}{d\xi}+\frac{d(nu)}{d\xi}=0, \label{eq1_41}
\end{equation}
\begin{equation}
-M\dfrac{du}{d\xi}+u\dfrac{du}{d\xi}=\dfrac{d\phi}{d\xi}, \label{eq1_42}
\end{equation}
\begin{equation}
\frac{d^{2}\phi}{d\xi^{2}}=-(1+\beta)+n+\beta\left(  1-\frac{\phi}
{\kappa-\tfrac{3}{2}}\right)  ^{-\kappa+1/2}. \label{eq1_43}
\end{equation}

Integration of the continuity equation and the equation of motion provides
\begin{equation}
\begin{array}
[c]{cc}
u=M(1-\dfrac{1}{n}),\text{  } & u=M-(M^{2}+2\phi)^{1/2}.
\end{array}
\label{eq1_44}
\end{equation}
Combining the above equations, we get
\begin{equation}
n=\left(  1+\frac{2\phi}{M^{2}}\right)  ^{-1/2}. \label{eq1_46}
\end{equation}
Substitution of the density expression (\ref{eq1_46}) into Poisson's equation
(\ref{eq1_43}) leads to
\begin{equation}
\frac{d^{2}\phi}{d\xi^{2}}=-\Psi_{1}^{\prime}(\phi,M,\beta,\kappa
)=-(1+\beta)+\left(  1+\frac{2\phi}{M^{2}}\right)  ^{-1/2}+\beta\left(
1-\frac{\phi}{\kappa-\tfrac{3}{2}}\right)  ^{-\kappa+1/2}, \label{eq1_47}
\end{equation}
where we use the definition $\Psi^{\prime}\equiv d\Psi/d\phi$  and
$\Psi^{\prime\prime}\equiv d^{2}\Psi/d\phi^{2}$ everywhere.

We impose the appropriate boundary conditions for localized waves: densities
are set to their unperturbed value at infinity, cold electron velocities and
the electrostatic potential are set to zero, i.e. $n=1$, $u=0$, and $\phi=0$.
The Poisson Eq. (\ref{eq1_43}) can be integrated to yield the energy integral,
\begin{equation}
\frac{1}{2}\left(  \frac{d\phi}{d\xi}\right)  ^{2}+\Psi_{1}(\phi
,M,\beta,\kappa)=0, \label{eq1_48}
\end{equation}
where $\Psi_{1}(\phi,M,\beta,\kappa)$ is the Sagdeev pseudopotential given by
\begin{align}
\Psi_{1}(\phi,M,\beta,\kappa)  &  =(1+\beta)\phi+M^{2}\left(  1-\left(
1+\frac{2\phi}{M^{2}}\right)  ^{1/2}\right) \nonumber\\
&  +\beta\left(  1-\left(  1+\frac{\phi}{-\kappa+\tfrac{3}{2}}\right)
^{-\kappa+3/2}\right)  . \label{eq1_49}
\end{align}
The Sagdeev pseudopotential depends on the Mach number $M$, the density ratio
$\beta$, and $\kappa$, and that $\Psi_{1}(\phi,M,\beta,\kappa)|_{\phi=0}=0$.
To obtain the electron-acoustic solitons, we must have an upper limit
$\phi=\phi_{\mathrm{m}}$, in which $\Psi_{1}(\phi,M,\beta,\kappa)|_{\phi
=\phi_{\mathrm{m}}}=0$. Here, we see that Eq. (\ref{eq1_48}) shows the form of
an energy balance equation. Accordingly, it can describe a motion of a
particle inside an anharmonic potential, i.e. the particle moves forward and
backward between the origin $\phi=0$ and the maximum position $\phi
=\phi_{\mathrm{m}}$. Obviously, Eq. (\ref{eq1_46}) is a real (non-imaginary)
expression for $\phi>-M^{2}/2$, so the maximum position for real solution is
given by $\phi_{\max}=-M^{2}/2$. A negative potential solitary wave may exist
if we can find a maximum peak of electrostatic wave potential $\phi
_{\mathrm{m}}$ ($<0$) by solving $\Psi_{1}(\phi,M,\beta,\kappa)=0$.

\subsection{Hot Electron Effect on EA Solitons}

\begin{figure}
\begin{center}
\includegraphics[width=6.0in]{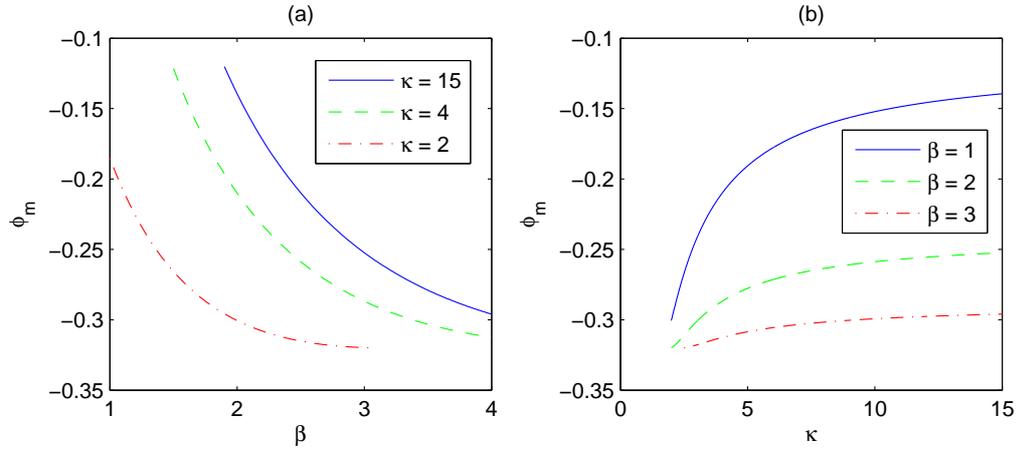}
\caption[Variation of $\phi_{\mathrm{m}}$ (a) with $\beta$ for different
values of  $\kappa$, and (b) with $\kappa$  for different values
of  $\beta$.]{Variation of $\phi_{\mathrm{m}}$: (a) with $\beta$ for different
values of  $\kappa$. Curves from bottom to top: $\kappa=2$ (solid),
$4$ (dashed), $15$ (dot-dashed). (b) with $\kappa$  for different values
of  $\beta$. Curves from top to bottom: $\beta=1$ (solid), $2$ (dashed),
$3$ (dot-dashed curve). Here, the Mach number is $0.8$.}
\label{fig1_4}
\end{center}
\end{figure}

Fig. \ref{fig1_4} shows the variation of the maximum electrostatic potential
$\phi_{\mathrm{m}}$ with $\beta$ for different values of  $\kappa$, and vice
versa. We can see that the absolute maximum electrostatic potential
$|\phi_{\mathrm{m}}|$ increases with either the rise in the ratio $\beta$ or
the decline in the parameter $\kappa$.

\begin{figure}
\begin{center}
\includegraphics[width=6.0in]{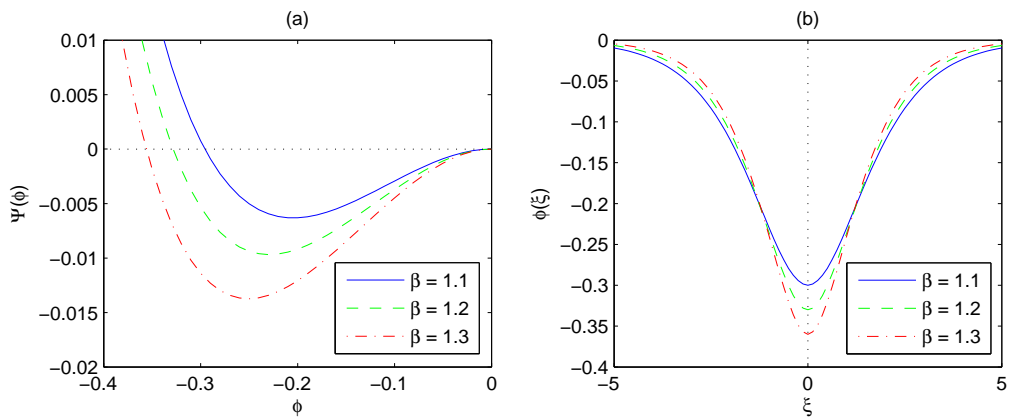}
\caption[(a) Variation of pseudopotential $\Psi(\phi)$ with $\phi$ for
different density ratio $\beta$. (b) Variation of potential $\phi$ with $\xi
$ for different density ratio $\beta$.]{(a) Variation of pseudopotential $\Psi(\phi)$ with $\phi$ for
different density ratio $\beta$. (b) Variation of potential $\phi$ with $\xi
$ for different density ratio $\beta$. Curves from top to bottom: $\beta
=1.1$ (solid), $1.2$ (dashed), $1.3$ (dot-dashed curve). Here, $\kappa
=3$ and $M=1$.}
\label{fig1_5}
\end{center}
\end{figure}

\begin{figure}
\begin{center}
\includegraphics[width=4.0in]{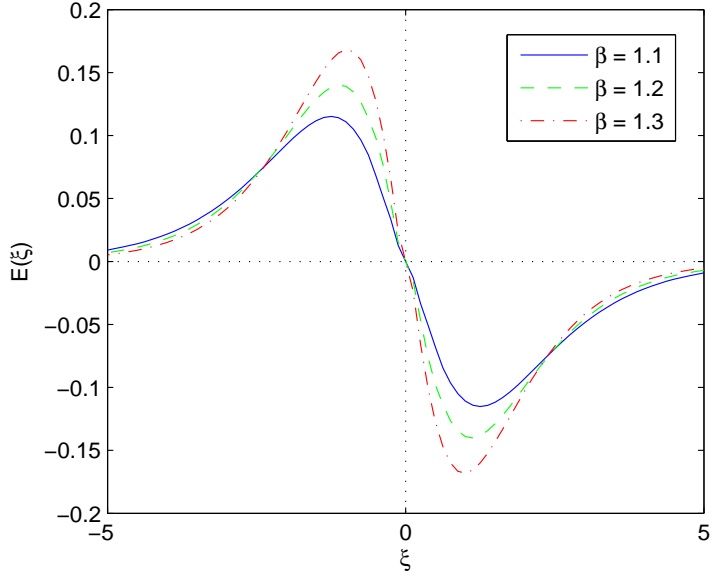}
\caption[Variation of electric field of the EAWs $E(\xi)$ with $\xi$ for
different density ratio $\beta$.]{Variation of electric field of the EAWs $E(\xi)$ with $\xi$ for
different density ratio $\beta$. Curves from bottom to top: $\beta
=1.1$ (solid), $1.2$ (dashed), $1.3$ (dot-dashed curve). Here, parameters
are same as used in Fig. \ref{fig1_5}.}
\label{fig1_6}
\end{center}
\end{figure}

In Fig. \ref{fig1_5}, it is seen that the electron-acoustic solitons have
negative perturbations of the electric potential. It shows the variation of
$\Psi(\phi)$ versus $\phi$ for different density ratio $\beta$. As the
density of the hot suprathermal electrons is increased, the potential
amplitude increases. In this case the associated electric field structures of
the EAWs are found to be bipolar, as shown in Fig. \ref{fig1_6} for different
value of  $\beta$. We can see that the increase in the number density of the
hot electrons raises the electric field's peak.

\section{Existence Conditions for Solitons}\label{singlecold:nonlinear:existence}

\begin{figure}
\begin{center}
\includegraphics[width=6.0in]{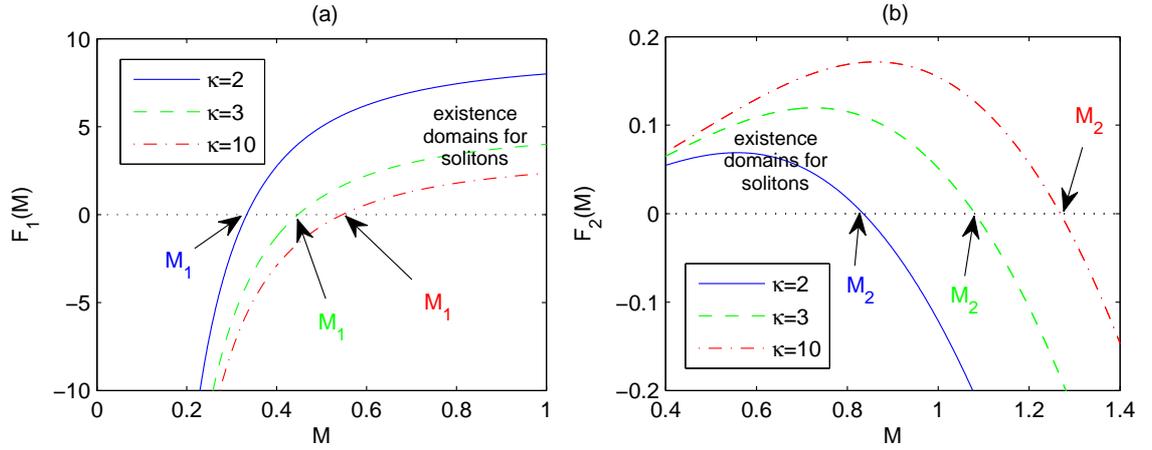}
\caption[The existence domains for stationary solitary structures.]{The existence domains for stationary solitary structures: (a) the
lower limit ($M_{1}$), (b) the upper limit ($M_{2}$). Curves from top to
bottom: $\kappa=2$ (solid), $3$ (dashed), $10$ (dot-dashed curve). Here,
$\beta=3$, and the quantities $F_{1}$ and $F_{2}$ are defined in
(\ref{eq1_53}) and (\ref{eq1_55}).}
\label{fig1_18}
\end{center}
\end{figure}

To obtain the electron-acoustic solitons, the conditions for the existence of
solitons, namely $\Psi_{1}^{\prime}(\phi,M,\beta,\kappa)=0$ and $\Psi
_{1}^{\prime\prime}(\phi,M,\beta,\kappa)<0$ at $\phi=0$, must be satisfied
(physically, $\phi=0$ is equilibrium; the potential $\Psi$ needs to have a
maximum, an unstable fixed point, at equilibrium; see Fig. \ref{fig1_5}a). The
lower limit for the Mach number is then obtained from the condition
\begin{equation}
F_{1}(M,\beta,\kappa)\equiv-\left.  \Psi_{1}^{\prime\prime}(\phi
,M,\beta,\kappa)\right\vert _{\phi=0}=\frac{\beta(\kappa-\frac{1}{2})}
{\kappa-\tfrac{3}{2}}-\frac{1}{M^{2}}>0 . \label{eq1_53}
\end{equation}
Eq. (\ref{eq1_53}) in terms of the Mach number defines a critical value as a
lower limit for $M$, i.e.
\begin{equation}
M_{1}(\beta,\kappa)=\left(  \frac{\kappa-\tfrac{3}{2}}{\beta(\kappa-\frac
{1}{2})}\right)  ^{1/2}. \label{eq1_54}
\end{equation}
Soliton solutions may exist only for values of the Mach number $M>M_{1}(\beta,\kappa)$ (lower limit). We notice that $M_{1}$ depends on the
parameters $\beta$ and $\kappa$. Figure \ref{fig1_18} (a) illustrates the
modification in the existence domains for different values of $\kappa$.

We obtain the largest possible value of $M$  through $\Psi_{1}(\phi
,M,\beta,\kappa)>0$ at $\phi=\phi_{\max}=-M^{2}/2$. This leads to the
following equation:
\begin{equation}
F_{2}(M,\beta,\kappa)\equiv M^{2}\left(  1-\dfrac{1}{2}(1+\beta)\right)
+\beta\left(  1-\left(  1+\frac{M^{2}}{2\kappa-3}\right)  ^{-\kappa
+3/2}\right)  >0. \label{eq1_55}
\end{equation}
The upper limit of the Mach number $M$ (say, $M_{2}(\beta,\kappa)$) is thus
obtained by solving the associated equation. As illustrated in Fig.
\ref{fig1_18} (b), the upper limit for the Mach number depends on the
parameter $\kappa$. From Eq. (\ref{eq1_55}), the upper limit ($M_{2}$) is
obviously modified by the density ratio $\beta$.

\begin{figure}
\begin{center}
\includegraphics[width=4.0in]{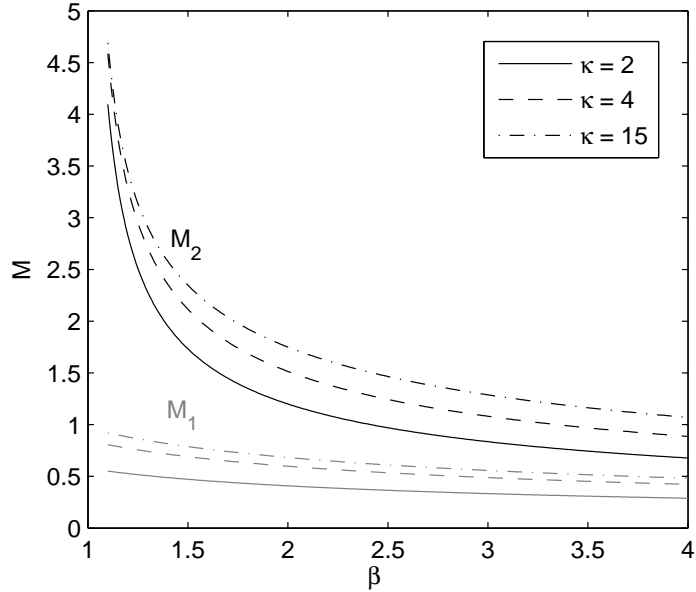}
\caption[Negative potential soliton existence domain in the parameter space of
$\beta$ and Mach number $M$.]{Negative potential soliton existence domain in the parameter space of
$\beta$ and Mach number $M$. Solitons may be supported in the region between
$M_{1}(\beta)$ (gray curve) and $M_{2}(\beta)$ (black curve). It shows
variation of $M_{1}(\beta)$ and $M_{2}(\beta)$ with $\beta$ for different
values of  $\kappa$. Curves from bottom to top: $\kappa=2$ (solid),
$4$ (dashed), $15$ (dot-dashed curve).}
\label{fig1_3}
\end{center}
\end{figure}

\subsection{Hot Suprathermal Effect on Velocity Range}

The existence domain is therefore derived from solving $F_{1}(M,\beta
,\kappa)>0$ and $F_{2}(M,\beta,\kappa)>0$. As illustrated in Fig.
\ref{fig1_18}, $M_{1}$ and $M_{2}$ increase with the increase in the parameter
$\kappa$. The range of Mach number ($M_{1}<M<M_{2}$) are shown in Fig.
\ref{fig1_3}, as function of equilibrium density ratio $\beta$ with the
various $\kappa$. As the density of the hot suprathermal electrons is
increased, the lower and upper limits of the Mach number decline. Hence, the
increase in the hot electrons causes the existence domain for stationary
solitary structure to become dramatically narrow. The minimum Mach number,
$M_{1}$, is generally less than the value of $1$. Especially, for the large
density ratio, $\beta>2.5$, the Maximum Mach number, $M_{2}$ becomes less than
$1.5$ as shown in Fig. \ref{fig1_3}.

\begin{figure}
\begin{center}
\includegraphics[width=6.0in]{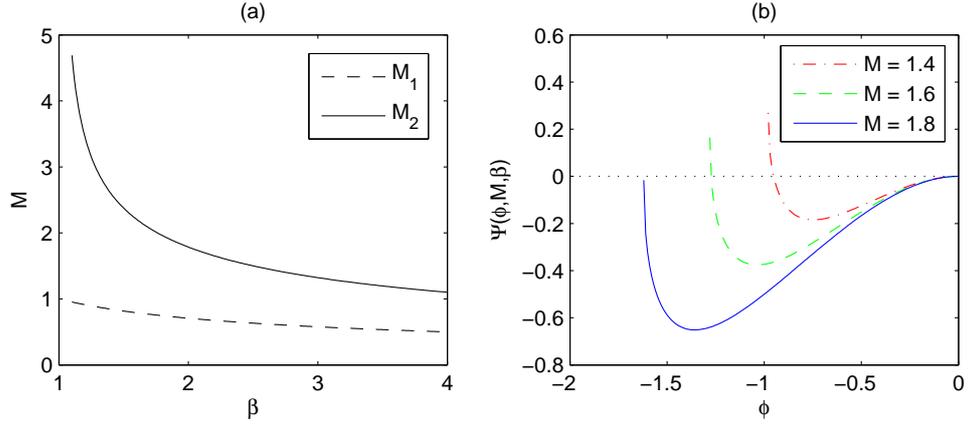}
\caption[EAWs in a plasma with hot Maxwellian electrons.]{EAWs in a plasma with hot Maxwellian electrons ($\kappa
\rightarrow\infty$). (a) Soliton existence domain in the parameter space of
$\beta$ and Mach number $M$. Solitons may be supported in the region between
$M_{1}(\beta)$ (dashed curve) and $M_{2}(\beta)$ (solid curve). (b)
Variation of $\Psi_{1}(\phi,M,\beta)$ for $\beta=2$ and different values of
Mach number, $M$. Curves from top to bottom: $M=1.2$ (dotted), $M=1.4$
 (dot-dashed), $M=1.6$ (dashed), and $M=1.8$ (solid).}
\label{fig1_2}
\end{center}
\end{figure}

\subsection{Velocity Range in Maxwellian vs. Suprathermal Plasmas}

\begin{figure}
\begin{center}
\includegraphics[width=4.0in]{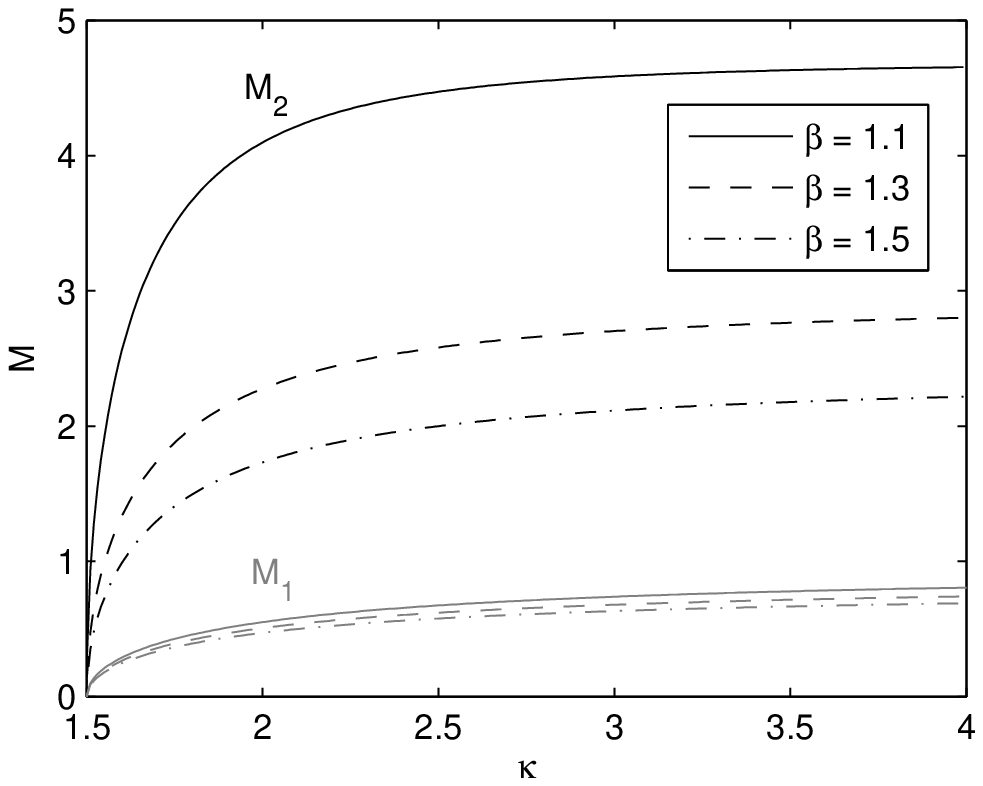}
\caption[Soliton existence domain in the parameter space of $\kappa$ and Mach
number $M$.]{Soliton existence domain in the parameter space of $\kappa$ and Mach
number $M$. Solitons may be supported in the region between $M_{1}$ (gray
curve) and $M_{2}$ (black curve). Curves from top to bottom: $\beta
=1.1$ (solid), $1.3$ (dashed), $1.5$ (dot-dashed curve). In the limit
$\kappa\rightarrow3/2$ easily see that $M_{1}=M_{2}=0$.}
\label{fig1_8}
\end{center}
\end{figure}

In the Maxwellian distributions for the hot electrons ($\kappa\rightarrow
\infty$), Eq. (\ref{eq1_53}) takes the following form
\begin{equation}
F_{1}(M,\beta)=\beta-\frac{1}{M^{2}}>0.
\end{equation}
This means that the lower limit becomes $M_{1}(\beta)=(\beta)^{-1/2}$. Eq.
(\ref{eq1_55}) tends to an exponential form
\begin{equation}
F_{2}(M,\beta)=M^{2}\left(  1-\dfrac{1}{2}(1+\beta)\right)  +\beta\left(
1-\exp(-\tfrac{1}{2}M^{2})\right)  >0. \label{eq1_56}
\end{equation}
The above equation solves the upper limit $M_{2}(\beta)$. Negative potential
solitary wave solutions of the cold electron fluid system of equations exist
for values of the Mach number $M$ in the range $M_{1}(\beta)<M<M_{2}(\beta)$,
which depends on the density ratios of the hot electrons to the cold
electrons. In Figure \ref{fig1_2} (a), we have plotted the lower and upper
limits, $M_{1}$ and $M_{2}$, respectively, over the range $1.1<\beta<4$ in the
limit $\kappa\rightarrow\infty$, and hence show the permitted range of Mach
numbers for the electron-acoustic solitons in the Maxwellian distributions. As
illustrated in Fig. \ref{fig1_2} (b) for the Maxwellian distributions, the
maximum electrostatic potential of the negative solitary structure increases
with the growth in the Mach number $M$ within the existence range
$M_{1}<M<M_{2}$. Furthermore, we can see that $M_{1}=M_{2}=0$ in the limit
$\kappa\rightarrow3/2$, as shown in Fig. \ref{fig1_8}. \allowbreak

%
%
%


\chapter{Warm Electron Fluid Model: Temperature Effects}\label{warmmodel}

In this chapter, we consider a collisionless and unmagnetized plasma
consisting of cool ($T_{c}\neq0$) inertial electrons, hot suprathermal
electron, and inertialess ions. We extend the thermal pressure to the model
described in \S  \ref{singlecold}. In \S  \ref{warmmodel:dr}, we obtain the
linear dispersion relation through using small deviations from the equilibrium
state. In \S  \ref{warmmodel:nonlinear}, we investigate the existence domain
of the electron-acoustic solitary waves.

The continuity equations of the cool electron fluid can be written as
\begin{equation}
\frac{\partial n_{c}}{\partial t}+\frac{\partial(n_{c}u_{c})}{\partial x}=0.
\label{eq2_1}
\end{equation}
Due to the thermal pressure of the cool electrons, the equation of momentum
contains an extra term (compare to Eq. (\ref{eq1_3}))
\begin{equation}
\frac{\partial u_{c}}{\partial t}+u_{c}\frac{\partial u_{c}}{\partial x}
=\frac{e}{m_{e}}\frac{\partial\phi}{\partial x}-\frac{1}{m_{e}n_{c}}
\frac{\partial P_{c}}{\partial x}. \label{eq2_2}
\end{equation}
The pressure of the cool electrons is given by
\begin{equation}
\frac{\partial P_{c}}{\partial t}+u_{c}\frac{\partial P_{c}}{\partial
x}+\gamma P_{c}\frac{\partial u_{c}}{\partial x}=0, \label{eq2_3}
\end{equation}
where $P_{c}$ is the thermal pressure of the cool electrons, $\gamma
=f+2/f\mathit{ }$denotes the specific heat ratio, and $f$ denotes the number
of degree of freedom, e.g., $\gamma=3$ in the one-dimensional case, also
$\gamma=1$ in an adiabatic evolution. We define the temperature ratio of the
cool electrons to the hot electrons as $\sigma=T_{c}/T_{h}$. The suprathermal
hot electron, fluid cool electrons and uniform ions are linked to the wave
potential by the Poisson's equation (\ref{eq1_4}).

The normalized one-dimensional ($\gamma=3$) model equations are written as
\begin{equation}
\frac{\partial n}{\partial t}+\frac{\partial(nu)}{\partial x}=0, \label{eq2_8}
\end{equation}
\begin{equation}
\frac{\partial u}{\partial t}+u\frac{\partial u}{\partial x}=\frac
{\partial\phi}{\partial x}-\frac{\sigma}{n}\frac{\partial P}{\partial x},
\label{eq2_9}
\end{equation}
\begin{equation}
\frac{\partial P}{\partial t}+u\frac{\partial P}{\partial x}+3P\frac{\partial
u}{\partial x}=0, \label{eq2_10}
\end{equation}
\begin{equation}
\frac{\partial^{2}\phi}{\partial x^{2}}=-(\beta+1)+n+\beta\left(  1-\frac
{\phi}{(\kappa-\tfrac{3}{2})}\right)  ^{-\kappa+1/2} . \label{eq2_11}
\end{equation}
The density $n_{c}$ are normalized with the unperturbed density ($n_{c,0}$),
the velocity $u_{c}$ with the hot electron thermal velocity ($c_{h,s}=\left(
k_{B}T_{h}/m_{e}\right)  ^{1/2}$), time with the inverse cool electron plasma
frequency, $\omega_{pc}^{-1}$, where $\omega_{pc}=(n_{c,0}e^{2}/\varepsilon
_{0}m_{e})^{1/2}$, length with the characteristic length scale, $\lambda
_{0}=(\varepsilon_{0}k_{B}T_{h}/n_{c,0}e^{2})^{1/2}$, the wave potential
$\phi$ with $k_{B}T_{h}/e$, and the thermal pressures with $n_{c,0}k_{B}T_{c}$.

\section{Dispersion Relation}\label{warmmodel:dr}

Let $S=(n,P,u,\phi)$ be any of the system variables describing the system's
state at a given position $x$ and instant $t$. We shall consider small
deviations from the equilibrium state $S^{(0)}=(1,1,0,0)$. Using  the
harmonic wave definition (\ref{eq1_15}), and the temporal and spatial
derivatives of the first order amplitudes, Eq. (\ref{eq1_17}), we get the
expressions for density, velocity, and pressure, namely
\begin{equation}
\begin{array}
[c]{ccc}
n_{1}^{(1)}=\dfrac{k}{\omega}u_{1}^{(1)},\text{  } & u_{1}^{(1)}=-\dfrac
{k}{\omega}\left(  \phi_{1}^{(1)}-\sigma P_{1}^{(1)}\right)  , & \text{
 }P_{1}^{(1)}=3n_{1}^{(1)}.
\end{array}
\label{eq2_17}
\end{equation}
The density in terms of potential are written as
\begin{equation}
n_{1}^{(1)}=-\left(  \dfrac{k^{2}}{\omega^{2}-3\sigma k^{2}}\right)  \phi
_{1}^{(1)}. \label{eq2_18}
\end{equation}

The system is closed by the Poisson's equation
\begin{equation}
-k^{2}\phi_{1}^{(1)}=-(\beta+1)+1+n_{1}^{(1)}+\beta\left(  1-\frac{\phi
_{1}^{(1)}}{\kappa-\tfrac{3}{2}}\right)  ^{-\kappa+1/2}. \label{eq2_19}
\end{equation}
Let us expand the $\kappa$-distribution as Eq. (\ref{eq1_20}), keeping up to
first order. Combining Eqs. (\ref{eq2_18}) and (\ref{eq2_19}), we get
\begin{equation}
-k^{2}\phi_{1}^{(1)}=-\left(  \dfrac{k^{2}}{\omega^{2}-3\sigma k^{2}}\right)
\phi_{1}^{(1)}+\beta\left(  \frac{\kappa-\frac{1}{2}}{\kappa-\tfrac{3}{2}
}\right)  \phi_{1}^{(1)}. \label{eq2_20}
\end{equation}
After a simplification, we recover the linear dispersion relation for the
electron-acoustic waves propagating in the warm model:
\begin{equation}
\omega_{2}^{2}=\frac{k^{2}}{k^{2}+k_{D}^{2}}+3{\sigma}k^{2}. \label{eq2_21}
\end{equation}
where $\sqrt{3{\sigma}}$ is the normalized thermal velocity. We note that
$\omega_{2}^{2}(k)=\omega_{1}^{2}(k)+3{\sigma}k^{2}$, where $\omega_{1}$ the
cold model frequency defined by Eq. (\ref{eq1_21}), and the warm model
frequency $\omega^{2}$ as in Eq. (\ref{eq2_21}).

Restoring dimensions, the warm model dispersion relation is derived as
\begin{equation}
\omega_{2}^{2}=\omega_{pc}^{2}\frac{k^{2}\lambda_{Dh}^{2}}{k^{2}\lambda
_{Dh}^{2}+\left(  \dfrac{\kappa-\frac{1}{2}}{\kappa-\tfrac{3}{2}}\right)
}+3\sigma k^{2}c_{h,s}^{2}. \label{eq2_23}
\end{equation}
For the limit $k\ll k_{D}$ Eq. (\ref{eq2_23}) reduces to $\omega\simeq
kv_{ph}$ where $v_{ph}$ is the phase speed given by
\begin{equation}
v_{ph}\simeq\left(  \beta^{-1}\left(  \frac{\kappa-\tfrac{3}{2}}{\kappa
-\frac{1}{2}}\right)  +3\sigma\right)  ^{1/2}c_{h,s}. \label{eq2_22}
\end{equation}
The thermal pressure manifests its physical effect in a small modification on
the $k$--$\omega$ plane. The linear dispersion relation is affected by the
thermal pressure.

\subsection{Temperature Effect on Linear Waves}

\begin{figure}
\begin{center}
\includegraphics[width=4.0in]{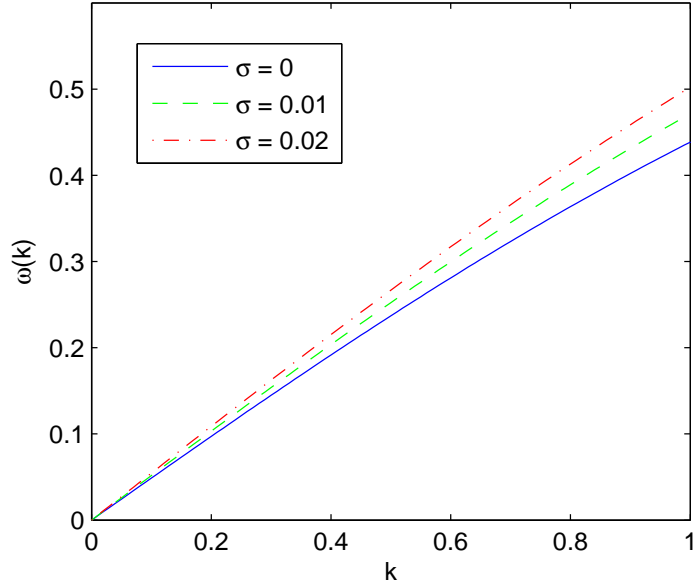}
\caption[Variation of the dispersion function curve for different values of $\sigma$]{Variation of the dispersion function curve for different values of
$\sigma$, $\beta=3$, and $\kappa=4$. Curves from bottom to top: $\sigma
=0$ (solid), $0.01$ (dashed), $0.02$ (dot-dashed curve).}
\label{fig2_2}
\end{center}
\end{figure}

Figure \ref{fig2_2} shows that the slope of the $\omega(k)$ curve increases
with a rise in the temperature ratio $\sigma$. Comparing Eqs. (\ref{eq1_23_2})
and (\ref{eq2_22}) we can see that growing $\sigma=T_{c}/T_{h}$ increases the
phase speed. It is obvious that in the limit $\sigma\rightarrow0$, Eq.
(\ref{eq1_21}), the cold model dispersion relation, is given.

\section{Sagdeev Pseudopotential Method}\label{warmmodel:nonlinear}

We take Eqs. (\ref{eq2_8})--(\ref{eq2_11}) to be stationary in a frame
traveling with velocity $M$ (the Mach number). Using the transformation
$\xi=x-Mt$, all temporal and spatial derivatives shall be replaced as Eq.
(\ref{eq1_40}), so Eqs. (\ref{eq2_8})--(\ref{eq2_11}) take the following form:
\begin{equation}
-M\dfrac{dn}{d\xi}+\frac{d(nu)}{d\xi}=0, \label{eq2_27}
\end{equation}
\begin{equation}
-M\dfrac{du}{d\xi}+u\dfrac{du}{d\xi}=\dfrac{d\phi}{d\xi}-\frac{\sigma}
{n}\dfrac{dP}{d\xi}, \label{eq2_28}
\end{equation}
\begin{equation}
-M\dfrac{dP}{d\xi}+u\dfrac{dP}{d\xi}+3P\dfrac{du}{d\xi}=0, \label{eq2_28_1}
\end{equation}
\begin{equation}
\frac{d^{2}\phi}{d\xi^{2}}=-(\beta+1)+n+\beta\left(  1-\frac{\phi}
{(\kappa-\tfrac{3}{2})}\right)  ^{-\kappa+1/2}. \label{eq2_29}
\end{equation}
Comparing Eqs. (\ref{eq2_27})--(\ref{eq2_29}) with Eqs.(\ref{eq1_41})--(\ref{eq1_43}), we see a thermal pressure in momentum equation that
classifies the propagation velocity as faster or slower than the electron thermal velocity.

Applying the appropriate boundary conditions, namely $n=1$, $P=1$, $u=0$, and
$\phi=0$, and integrating the equation of continuity, the equation of motion,
and the equation of state provide
\begin{equation}
\begin{array}
[c]{cc}
u=M(1-\dfrac{1}{n}),\text{  } & u={M-(M}^{2}{+2\phi-3n^{2}\sigma+3\sigma
)}^{1/2},
\end{array}
\label{eq2_30}
\end{equation}
\begin{equation}
P=n^{3}\rightarrow dP=3n^{2}dn. \label{eq2_31}
\end{equation}
Combining Eqs. (\ref{eq2_30}) and (\ref{eq2_31}), we obtain the following
solutions through the biquadratic equation (see Appendix
\ref{BiquadraticEquation} for more detail):
\begin{equation}
{n=}\dfrac{1}{2}\left(  n_{(+)}\pm n_{(-)}\right),
\label{eq2_32}
\end{equation}
\begin{equation}
\begin{array}
[c]{cc}
n_{(+)}{\equiv}\left(  \dfrac{{2\phi+}\left(  {M+}\sqrt{3{\sigma}}\right)
^{2}}{3{\sigma}}\right)  ^{1/2},\text{  } & n_{(-)}{\equiv}\left(
\dfrac{{2\phi+\left(  {M-}\sqrt{3{\sigma}}\right)  ^{2}}}{3{\sigma}}\right)
^{1/2}.
\end{array}
\end{equation}
In Eq. (\ref{eq2_32}), the upper sign ($+$) is for subsonic cool electrons
(${M<}\sqrt{3{\sigma}}$) soliton while the lower sign ($-$) is for supersonic
cool electrons (${M>}\sqrt{3{\sigma}}$), because it must satisfy the condition
at equilibrium ($n=1$ at $\phi=0$). We notice that the normalized density has
two regions in the Mach number domain, namely subsonic and supersonic for hot
species and cool species, respectively. We obtain the condition at equilibrium
($n=1$) at $\phi=0$. In the limit $\sigma\rightarrow0$, we recover the cold
limit expression (\ref{eq1_46}). To have the real solution, ${2\phi+\left(
{M-}\sqrt{3{\sigma}}\right)  ^{2}>0}$, so it yields $\phi_{\max}=-\frac{1}{2}\left(  {M-}\sqrt{3{\sigma}}\right)  ^{2}$ to the negative solitary
structures.

Substituting the density expression (\ref{eq2_32}) into the Poisson's equation
(\ref{eq2_29}) leads to the equation of motion:
\begin{align}
\frac{d^{2}\phi}{d\xi^{2}}  &  =-\Psi_{2}^{\prime}(\phi,M,\beta,\kappa
,{\sigma})=-(\beta+1)+\beta\left(  1-\frac{\phi}{(\kappa-\tfrac{3}{2}
)}\right)  ^{-\kappa+1/2}\nonumber\\
&  +\dfrac{1}{2\sqrt{3{\sigma}}}\left(  \left[  {2\phi+}\left(  {M+}
\sqrt{3{\sigma}}\right)  ^{2}\right]  ^{1/2}\pm\left[  {2\phi+\left(
{M-}\sqrt{3{\sigma}}\right)  ^{2}}\right]  ^{1/2}\right)  . \label{eq2_36}
\end{align}
The above equation can be integrated to yield the energy balance equation:
\begin{equation}
\frac{1}{2}\left(  \frac{d\phi}{d\xi}\right)  ^{2}+\Psi_{2}(\phi
,M,\beta,\kappa,{\sigma})=0, \label{eq2_37}
\end{equation}
where the Sagdeev pseudopotential $\Psi_{1}(\phi,M,\beta,\kappa,{\sigma})$
reads as
\begin{align}
\Psi_{2}(\phi,M,\beta,\kappa,{\sigma})  &  =(1+\beta)\phi+\beta\left(
1-\left(  1+\frac{\phi}{-\kappa+\tfrac{3}{2}}\right)  ^{-\kappa+3/2}\right)
\nonumber\\
&  +\frac{1}{6\sqrt{3{\sigma}}}\left(  \left(  {M+}\sqrt{3{\sigma}}\right)
^{3}\pm{{\left(  {M-}\sqrt{3{\sigma}}\right)  ^{3}}}\right. \nonumber\\
&  -\left.  \left[  {2\phi+}\left(  {M+}\sqrt{3{\sigma}}\right)  ^{2}\right]
^{3/2}\mp{\left[  {2\phi+\left(  {M-}\sqrt{3{\sigma}}\right)  ^{2}}\right]
}^{3/2}\right)  . \label{eq2_38}
\end{align}
Here, the upper sign is for subsonic soliton and the lower sign for
supersonic. It is easily seen that we get the cold model in the limit
$\sigma\rightarrow0$, i.e., $\lim_{\sigma\rightarrow0}\Psi_{2}(\phi
,M,\beta,\kappa,{\sigma})=\Psi_{1}(\phi,M,\beta,\kappa)$.

\subsection{Temperature Effect on EAWs}

\begin{figure}
\begin{center}
\includegraphics[width=6.0in]{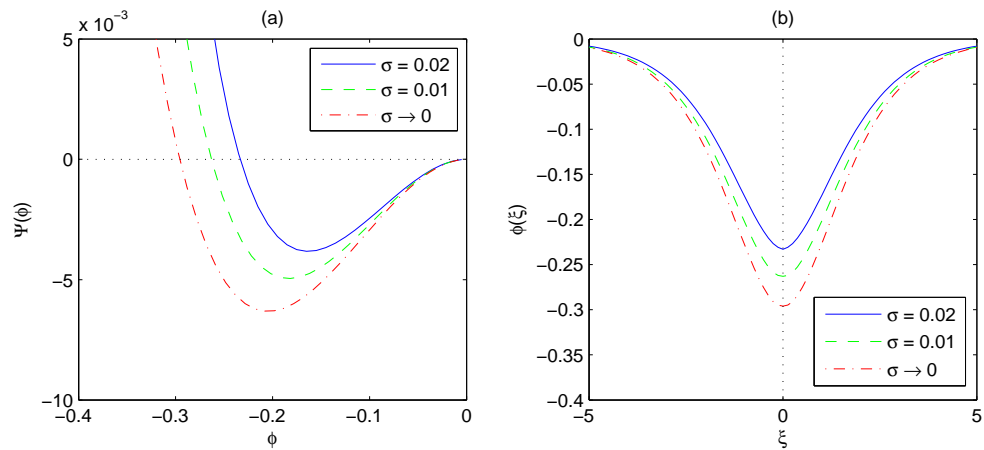}
\caption[(a) Variation of pseudopotential $\Psi(\phi)$ with $\phi$ for
different temperature ratio $\sigma$. (b) Variation of the electron-acoustic
potential $\phi$ with $\xi$ for different temperature ratio $\sigma$.]{(a) Variation of pseudopotential $\Psi(\phi)$ with $\phi$ for
different temperature ratio $\sigma$. (b) Variation of the electron-acoustic
potential $\phi$ with $\xi$ for different temperature ratio $\sigma$. Curves
from bottom to top: $\sigma\rightarrow0$ (dot-dashed curve), $\sigma
=0.01$ (dashed), $0.02$ (solid). Here, $\beta=1.1$, $\kappa=3$ and $M=1$.}
\label{fig2_5}
\end{center}
\end{figure}

\begin{figure}
\begin{center}
\includegraphics[width=5.0in]{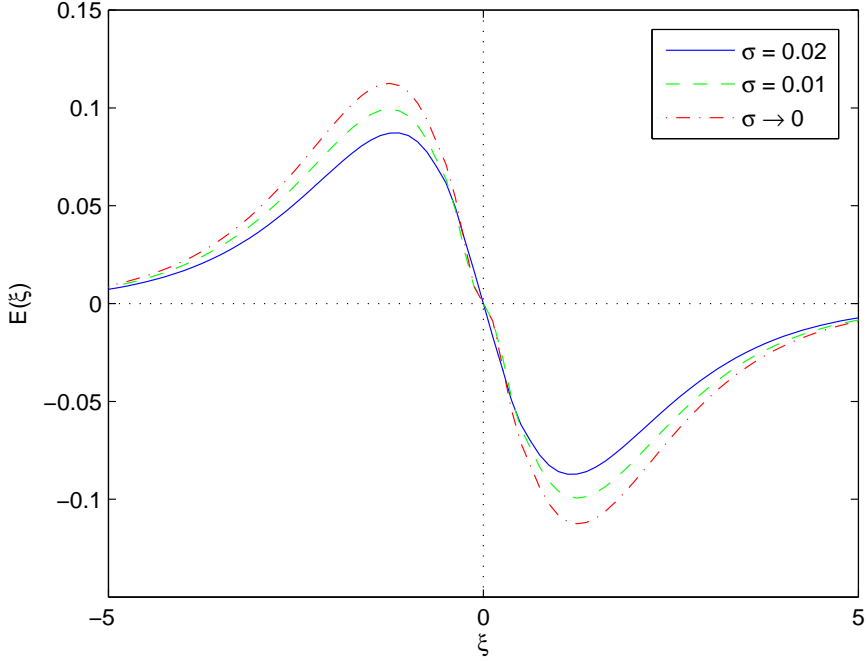}
\caption[Variation of electric field of the EAWs $E(\xi)$ with $\xi$ for
different temperature ratio $\sigma$.]{Variation of electric field of the EAWs $E(\xi)$ with $\xi$ for
different temperature ratio $\sigma$. Curves from top to bottom:
$\sigma\rightarrow0$ (dot-dashed curve), $\sigma=0.01$ (dashed),
$0.02$ (solid). Here, parameters are same as used in Fig. \ref{fig2_5}.}
\label{fig2_7}
\end{center}
\end{figure}

\begin{figure}
\begin{center}
\includegraphics[width=6.0in]{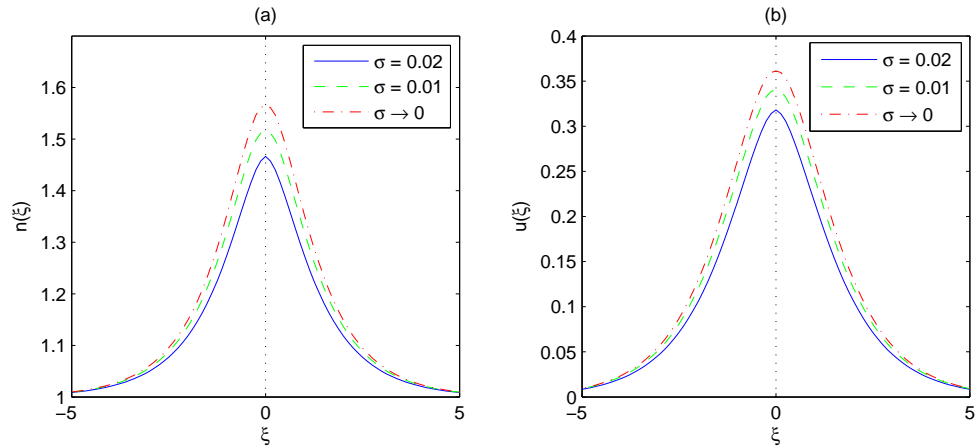}
\caption[(a) Variation of density $n$ with $\xi$ for different temperature
ratio $\sigma$. (b) Variation of velocity $u$ with $\xi$ for different
temperature ratio $\sigma$.]{(a) Variation of density $n$ with $\xi$ for different temperature
ratio $\sigma$. (b) Variation of velocity $u$ with $\xi$ for different
temperature ratio $\sigma$. Curves from top to bottom: $\sigma\rightarrow
0$ (dot-dashed curve), $\sigma=0.01$ (dashed), $0.02$ (solid). Here,
parameters are same as used in Fig. \ref{fig2_5}.}
\label{fig2_6}
\end{center}
\end{figure}

We have numerically solved Eq. (\ref{eq2_38}) for a plasma which consists of
cool electrons and hot suprathermal electrons. Figure \ref{fig2_5} (a) shows
the variation of Sagdeev pseudopotential $\Psi_{2}(\phi)$ with normalized
potential for different temperature ratio $\sigma$. Figure \ref{fig2_5} (b)
shows the variation of solitary waves for the cool electrons for different
values of the temperature ratio $\sigma=T_{c}/T_{h}$ as shown on the curves
for $\beta=1.1$, $\kappa=3$ and Mach number, $M=1$. The amplitude of the wave
potential decreases with the increase in $\sigma$. The associated bipolar
electric field structures are shown in Fig. \ref{fig2_7}. We can see a decline
in the electric field structures with an increase in the thermal velocity
$\sqrt{3\sigma}$. As illustrated in Fig. \ref{fig2_6}, the number density and
the velocity of the cool electrons decline with the growth in the thermal velocity.

\begin{figure}
\begin{center}
\includegraphics[width=6.0in]{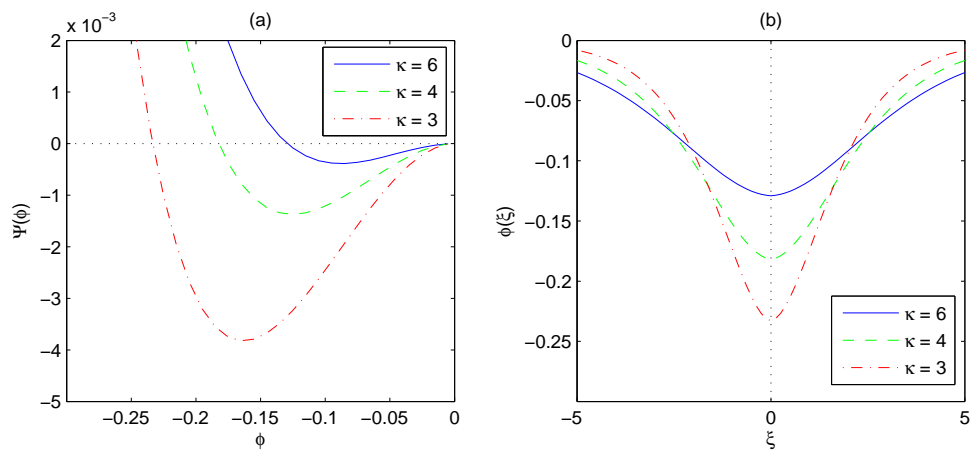}
\caption[(a) Variation of pseudopotential $\Psi(\phi)$ with $\phi$ for
different $\kappa$. (b) Variation of potential $\phi$ with $\xi$ for
different $\kappa$.]{(a) Variation of pseudopotential $\Psi(\phi)$ with $\phi$ for
different $\kappa$. (b) Variation of potential $\phi$ with $\xi$ for
different $\kappa$. Curves from bottom to top: $\kappa=6$ (solid),
$4$ (dashed), $3$ (dot-dashed curve). Here, $\sigma=0.02$, $\beta=1.1$, and
$M=1$.}
\label{fig2_8}
\end{center}
\end{figure}

\begin{figure}
\begin{center}
\includegraphics[width=5.0in]{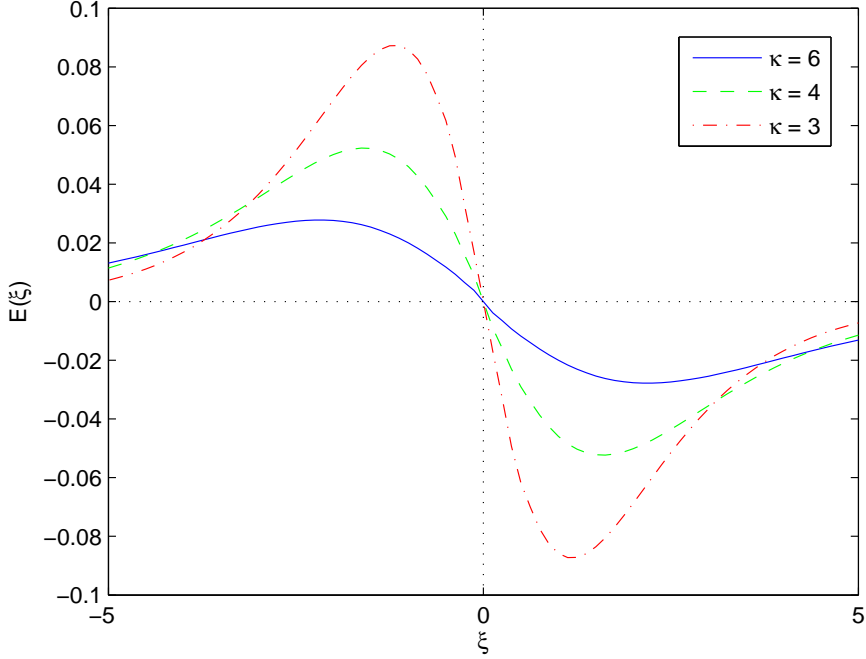}
\caption[Variation of electric field of the EAWs $E(\xi)$ with $\xi$ for
different $\kappa$.]{Variation of electric field of the EAWs $E(\xi)$ with $\xi$ for
different $\kappa$. Curves from top to bottom: $\kappa=6$ (solid),
$4$ (dashed), $3$ (dot-dashed curve). Here, parameters are same as used in
Fig. \ref{fig2_8}.}
\label{fig2_10}
\end{center}
\end{figure}

\begin{figure}
\begin{center}
\includegraphics[width=6.0in]{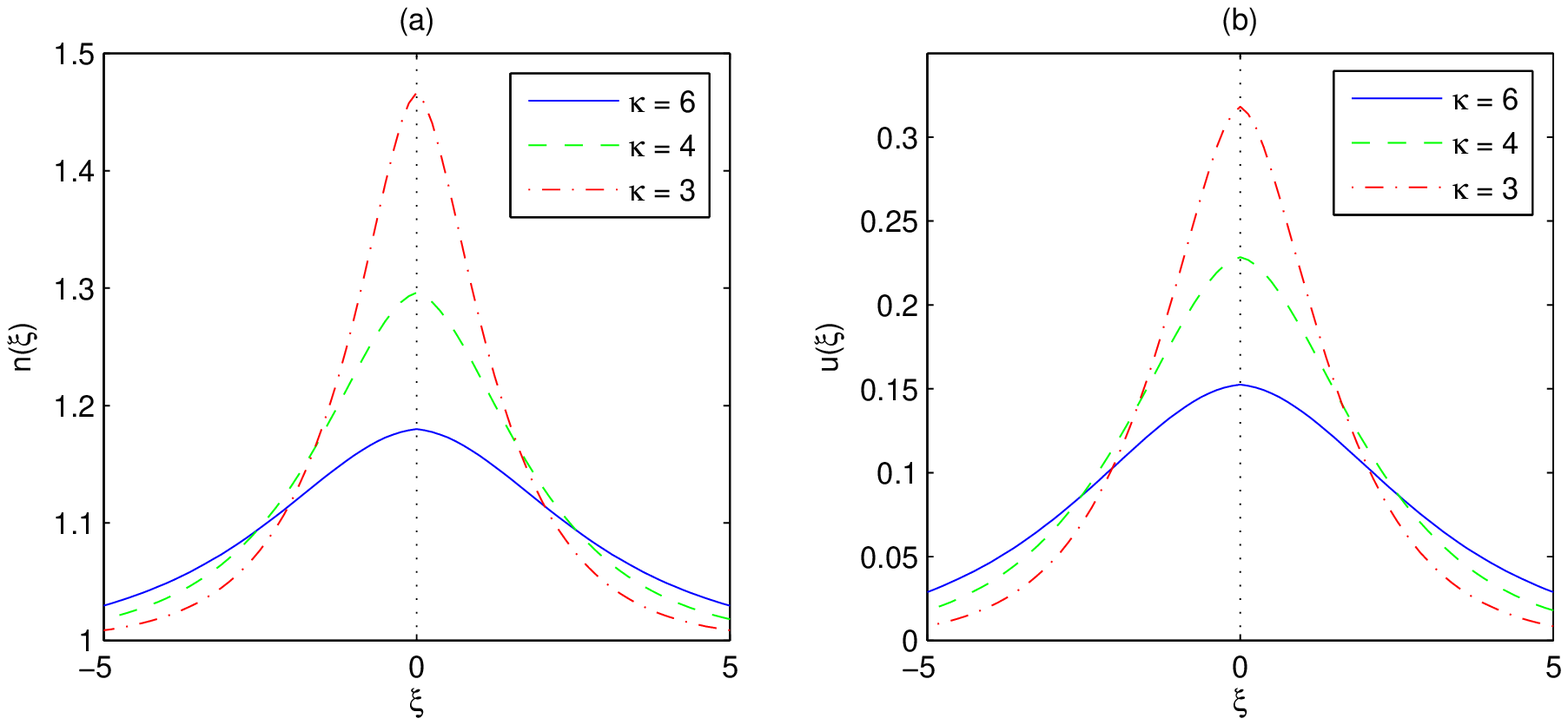}
\caption[(a) Variation of density $n$ with $\xi$ for different $\kappa$. (b)
Variation of velocity $u$ with $\xi$ for different $\kappa$.]{(a) Variation of density $n$ with $\xi$ for different $\kappa$. (b)
Variation of velocity $u$ with $\xi$ for different $\kappa$. Curves from top
to bottom: $\kappa=6$ (solid), $4$ (dashed), $3$ (dot-dashed curve). Here,
parameters are same as used in Fig. \ref{fig2_8}.}
\label{fig2_9}
\end{center}
\end{figure}

\subsection{Suprathermal Effect on EAWs}

Figure \ref{fig2_8} (a) shows the variation of Sagdeev pseudopotential
$\Psi_{2}(\phi)$ versus $\phi$ for different $\kappa$. The absolute maximum
electrostatic potential $|\phi_{\mathrm{m}}|$ decrease with the rise in
$\kappa$, while the large $\kappa$ turns into Maxwellian distribution. The
value of $\kappa$ between $3/2$ and $6$ effectively describe the solitary
structure of the electron-acoustic wave in a suprathermal plasma. Figure
\ref{fig2_10} shows the variation of the associated bipolar electric field
structures for different values of $\kappa$. In Fig. \ref{fig2_9}, we can see
the density $n$ and the velocity $u$ increase, as the parameter $\kappa$ is decreased.

\section{Soliton Existence}\label{warmmodel:nonlinear:existence}

We require to find out if the conditions for the existence of solitons are
satisfied for Eq. (\ref{eq2_38}), i.e., $\Psi_{2}^{\prime}(\phi,M,\beta
,\kappa,{\sigma})=0$ and $\Psi_{2}^{\prime\prime}(\phi,M,\beta,\kappa,{\sigma
})<0$ at $\phi=0$. We derive the lower limit for the existence domain from the
condition
\begin{equation}
F_{1}(M,\beta,\kappa,{\sigma})=-\left.  \Psi_{2}^{\prime\prime}(\phi
,M,\beta,\kappa,{\sigma})\right\vert _{\phi=0}=\frac{\beta(\kappa-\frac{1}
{2})}{\kappa-\tfrac{3}{2}}-\frac{1}{M^{2}-3{\sigma}}>0. \label{eq2_44}
\end{equation}
Eq. (\ref{eq2_44}) provides the minimum value for the Mach number:
\begin{equation}
M_{1}(\beta,\kappa,{\sigma})=\left(  \frac{\kappa-\tfrac{3}{2}}{\beta
(\kappa-\frac{1}{2})}+3{\sigma}\right)  ^{1/2}. \label{eq2_45}
\end{equation}
Soliton solutions may exist only for the Mach number greater than $M_{1}
(\beta,\kappa,{\sigma})$. We can see that $M_{1}$ depends on the parameters
$\beta$, $\kappa$, and $\sigma$. This shows that electron thermal effects
increase the Mach number threshold. In the limit $\sigma\rightarrow0$, it
provides the expression for cold model (\ref{eq1_54}).

\begin{figure}
\begin{center}
\includegraphics[width=6.0in]{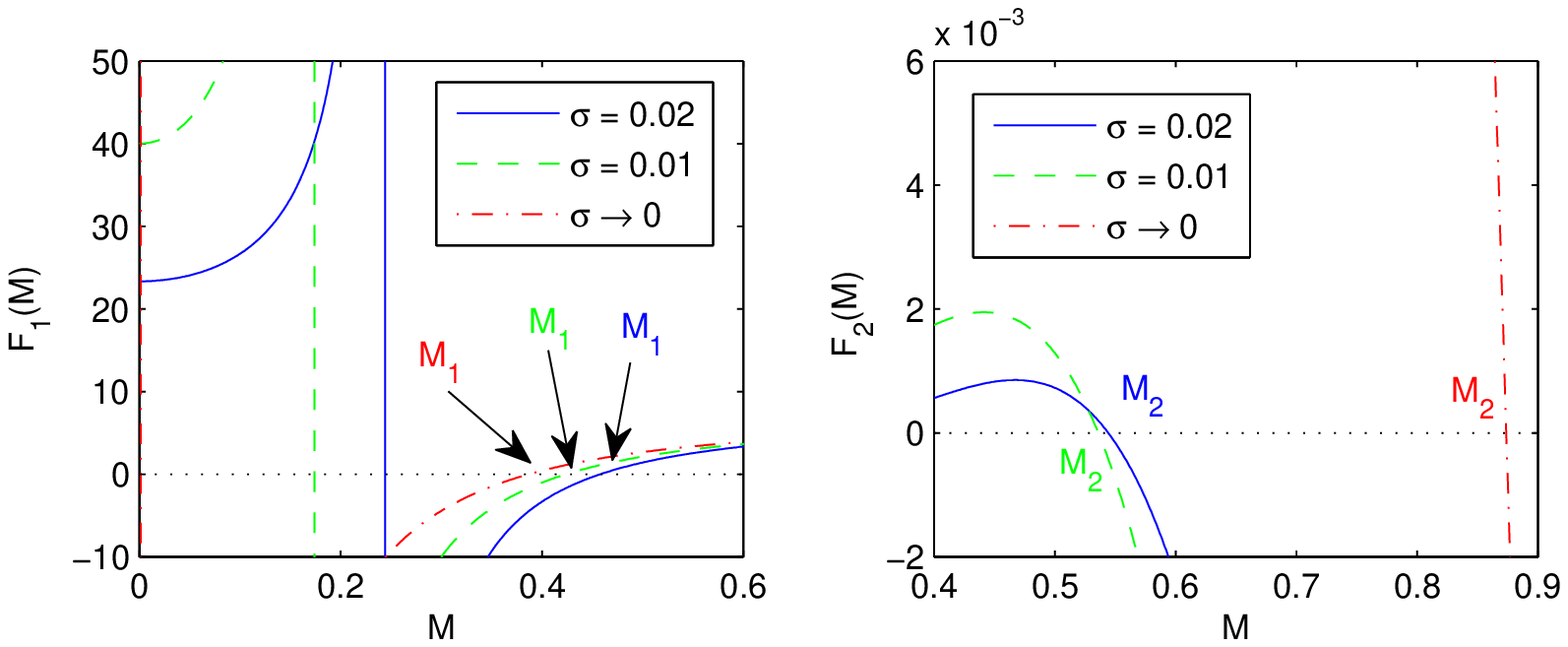}
\caption[The existence domains for stationary solitary structures.]{The existence domains for stationary solitary structures: (a) the
lower limit ($M_{1}$), (b) the upper limit ($M_{2}$). Curves: $\sigma
\rightarrow0$ (dot-dashed curve), $\sigma=0.01$ (dashed), $0.02$ (solid).
Here, $\beta=4$ and $\kappa=3$. The quantities $F_{1}$ and $F_{2}$ are
defined in (\ref{eq2_44}) and (\ref{eq2_47})}
\label{fig2_18}
\end{center}
\end{figure}

We obtain the largest possible value of $M$  through $F_{2}(M,\beta
,\kappa,{\sigma}) $ = $\Psi_{2}(\phi,M,$ $\beta,\kappa,{\sigma})|_{\phi=\phi_{\max}}>0$. This yields the following equation:
\begin{align}
F_{2}(M,\beta,\kappa,{\sigma})  &  =-\frac{1}{2}(1+\beta)\left(  {M-}
\sqrt{3{\sigma}}\right)  ^{2}+M{{^{2}+}\sigma}-\frac{4}{3}M\sqrt
{M\sqrt{3{\sigma}}}\nonumber\\
&  +\beta\left(  1-\left(  1+\frac{\left(  {M-}\sqrt{3{\sigma}}\right)  ^{2}
}{2\kappa-3}\right)  ^{-\kappa+3/2}\right)  . \label{eq2_47}
\end{align}
Solving Eq. (\ref{eq2_47}) provides the upper limit $M_{2}(\alpha,\kappa)$ for
the Mach number. Figure \ref{fig2_18} illustrates a modification in the
existence domains ($M_{1}<M<M_{2}$) for different values of $\sigma$. We find
out that \textquotedblleft cool\textquotedblright electrons need to be
supersonic (in the sense ${M>}\sqrt{3{\sigma}}$) and \textquotedblleft
hot\textquotedblright suprathermal electrons subsonic (${M<}\sqrt{3{\sigma}}$) \cite{Verheest2004,McKenzie2004,Verheest2007}. Negative solitary structures
of the cool electron-fluid may be found in the range $M_{1}<M<M_{2}$, which
depends on the parameters $\beta$, $\kappa$, and $\sigma$.

\subsection{Velocity Range in Maxwellian vs. Suprathermal Plasmas}

\begin{figure}
\begin{center}
\includegraphics[width=6.0in]{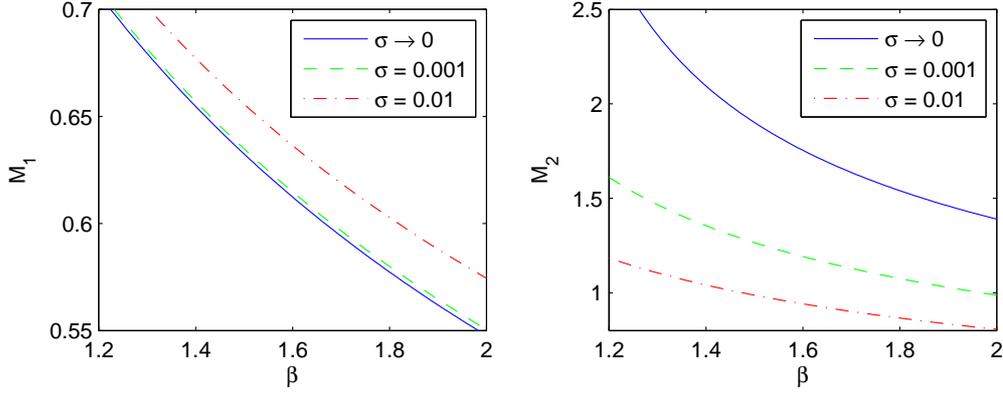}
\caption[Variation of the lower limit $M_{1}$ and the upper limit $M_{2}$
with $\beta$ for different temperature ratio $\sigma$.]{Variation of the lower limit $M_{1}$ and the upper limit $M_{2}$
with $\beta$ for different temperature ratio $\sigma$. Curves from bottom
to top: $\sigma\rightarrow0$ (solid), $\sigma=0.001$ (dashed),
$0.01$ (dot-dashed curve). Here, $\kappa=3$.}
\label{fig2_11}
\end{center}
\end{figure}

\begin{figure}
\begin{center}
\includegraphics[width=6.0in]{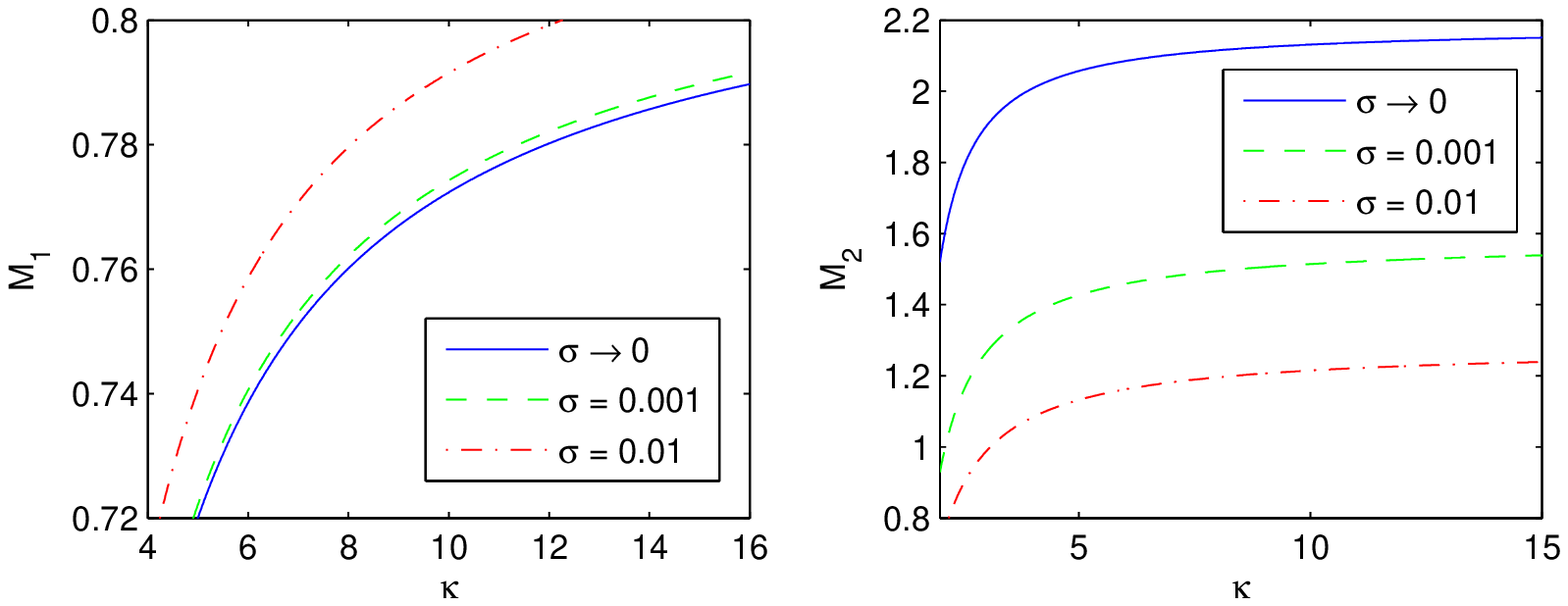}
\caption[Variation of the lower limit $M_{1}$ and the upper limit $M_{2}
$ with $\kappa$ for different temperature ratio $\sigma$.]{Variation of the lower limit $M_{1}$ and the upper limit $M_{2}
$ with $\kappa$ for different temperature ratio $\sigma$. Curves from
bottom to top: $\sigma\rightarrow0$ (solid), $\sigma=0.01$ (dashed),
$0.001$ (dot-dashed curve). Here, $\beta=1.5$.}
\label{fig2_13}
\end{center}
\end{figure}

\begin{figure}
\begin{center}
\includegraphics[width=5.0in]{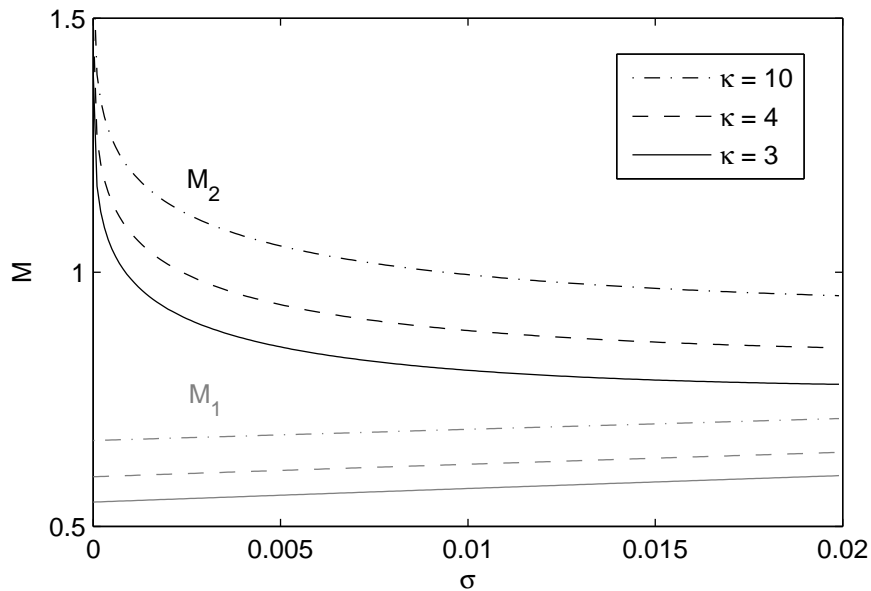}
\caption[Negative potential soliton existence domain in the parameter space of
$\sigma$ and Mach number $M$.]{Negative potential soliton existence domain in the parameter space of
$\sigma$ and Mach number $M$. Solitons may be supported in the region between
$M_{1}(\sigma)$ (gray curve) and $M_{2}(\sigma)$ (black curve). It shows
variation of $M_{1}(\sigma)$ and $M_{2}(\sigma)$ with $\sigma$. Curves from
bottom to top: $\kappa=3$ (solid), $4$ (dashed), $10$ (dot-dashed curve).
Here, $\beta=2$.}
\label{fig2_15}
\end{center}
\end{figure}

In the Maxwellian distributions ($\kappa\rightarrow\infty$), we get
\begin{equation}
F_{1}(M,\beta,{\sigma})=\beta-\frac{1}{M^{2}-3{\sigma}}>0.
\end{equation}
\begin{align}
F_{2}(M,\beta,{\sigma})  &  =-\frac{1}{2}(1+\beta)\left(  {M-}\sqrt{3{\sigma}
}\right)  ^{2}+M{{^{2}+}\sigma}-\frac{4}{3}M^{3/2}\left(  3{\sigma}\right)
^{1/4}\nonumber\\
&  +\beta\left(  1-\exp(-\tfrac{1}{2}\left(  {M-}\sqrt{3{\sigma}}\right)
^{2})\right)  .
\end{align}
The above equation solves the upper limit $M_{2}(\beta)$, while the lower
limit becomes $M_{1}(\beta,{\sigma})=\left(  1/\beta+3{\sigma}\right)  ^{1/2}
$. As shown in Figures \ref{fig2_11}--\ref{fig2_15}, growing the thermal
pressure pushes up the lower limit $M_{1}$, but turns down the upper limit
$M_{2}$ of the Mach number. We can also see the decline in both $M_{1}$ and
$M_{2}$ with the increase in $\beta$ and decrease in $\kappa$, which has been
previously described in \S  \ref{singlecold:nonlinear:existence}.

In the limit $\kappa\rightarrow3/2$, the lower limit of the Mach number takes
the form $M_{1}(\beta,{\sigma})=\sqrt{3{\sigma}}$. It is non-zero, in contrast
to the cold model in \S  \ref{singlecold:nonlinear:existence} which turned
into zero. The upper limit $M_{2}$ can be solved by
\begin{equation}
F_{2}(M,\beta,{\sigma})=-\frac{1}{2}(1+\beta)\left(  {M-}\sqrt{3{\sigma}
}\right)  ^{2}+M{{^{2}+}\sigma}-\frac{4}{3}M^{3/2}\left(  3{\sigma}\right)
^{1/4}>0.
\end{equation}
It also appears to be, nonvanishing, in proportion to the thermal velocity,
$M_{2}\sim\sqrt{3{\sigma}}$. In the limit $\sigma\rightarrow0$, we obtain the
cold model results ($M_{1}=M_{2}=0$).

\subsection{Temperature Effect on Velocity Range}

The existence condition ($M_{1}<M<M_{2}$) is obtained through $F_{1}
(M,\beta,\kappa,{\sigma})>0$ and $F_{2}(M,\beta,\kappa,{\sigma})>0$. Fig.
\ref{fig2_11} shows that $M_{1}$ and $M_{2}$ decline with the increase in the
parameter $\beta$, i.e., the density of the hot electrons. We notice the
existence domain becomes narrower, as the hot electrons density is increased.
The range of the Mach number are shown in Fig. \ref{fig2_13}, as function of
$\kappa$ with the various $\sigma$. In this figure, one can see that, moving
into the Maxwellian distribution ($\kappa\rightarrow\infty$) will broaden the
Mach number range. However, the lower Mach number limit tend to $\sqrt
{3{\sigma}}$, and the upper Mach number limit to $\sqrt{3{\sigma}}$ as
$\kappa\rightarrow3/2$, the limiting value of $\kappa$. As illustrated in Fig.
\ref{fig2_15} for suprathermal situation ($3/2<\kappa<6$), the lower Mach
number limit, $M_{1}$, is generally less than the value of $0.75$, and the
upper Mach number limit, $M_{2}$, for very warm model ($\sigma>0.005$) becomes
less than $1$.

%
%
%


\chapter{Two-Fluid Model: Ion Inertia Effects}\label{twofluidmodel}

In this chapter, we consider a collisionless and unmagnetized plasma with
three components, namely, cool inertial electrons, inertialess hot
suprathermal electrons, and inertial ions. We include the inertial ions in the
model described in Chapter \ref{warmmodel}. We employ the cool electrons
described by Eqs. (\ref{eq2_1})--(\ref{eq2_3}), the hot suprathermal
electrons, assumed to obey the kappa velocity distribution (\ref{eq1_1}), and
ions, described by the fluid-moment equations. The electron-fluid and
ion-fluid are coupled through Poisson's equation (\ref{eq1_4}). In
\S  \ref{twofluidmodel:dr}, we derive the linear dispersion relation from a
linear methodology. In \S  \ref{twofluidmodel:nonlinear}, we develop a
Sagdeev pseudopotential method and determine the existence domain of
stationary solitary waves.

The fluid equations for the ions read
\begin{equation}
\frac{\partial n_{i}}{\partial t}+\frac{\partial(n_{i}u_{i})}{\partial x}=0,
\label{eq3_4}
\end{equation}
\begin{equation}
\frac{\partial u_{i}}{\partial t}+u_{i}\frac{\partial u_{i}}{\partial
x}=-\frac{Ze}{m_{i}}\frac{\partial\phi}{\partial x}, \label{eq3_5}
\end{equation}
where $n_{i,0}$ is the density of the ions in the undisturbed plasma, $m_{e}$
the mass of the ions, $Z$ the number of ions (everywhere, $Z=1$).

The normalized fluid-moment equations of the cool electron and the ions, and
the Poisson's equation are written as Eqs. (\ref{eq2_8})--(\ref{eq2_10}), and
\begin{equation}
\frac{\partial\tilde{n}}{\partial t}+\frac{\partial(\tilde{n}\tilde{u}
)}{\partial x}=0, \label{eq3_12}
\end{equation}
\begin{equation}
\frac{\partial\tilde{u}}{\partial t}+\tilde{u}\frac{\partial\tilde{u}
}{\partial x}=-\mu\frac{\partial\phi}{\partial x}, \label{eq3_13}
\end{equation}
\begin{equation}
\frac{\partial^{2}\phi}{\partial x^{2}}=-\tilde{n}+n+\beta\left(  1-\frac
{\phi}{(\kappa-\tfrac{3}{2})}\right)  ^{-\kappa+1/2}, \label{eq3_14}
\end{equation}
All densities are normalized with the unperturbed density of the cool
electrons ($n_{c,0}$), all velocities with the hot electron thermal velocity
($c_{h,s}=\sqrt{k_{B}T_{h}/m_{e}}$):
\begin{equation}
\begin{array}
[c]{cccc}
\dfrac{n_{c}}{n_{c,0}}\rightarrow n,\text{  } & \dfrac{n_{i}}{n_{c,0}
}\rightarrow\tilde{n},\text{  } & \dfrac{u_{c}}{c_{h,s}}\rightarrow u,\text{
 } & \dfrac{u_{i}}{c_{h,s}}\rightarrow\tilde{u},
\end{array}
\label{eq3_15}
\end{equation}
space and time variables are scaled by the characteristic length scale,
$\lambda_{0}=\left(  \varepsilon_{0}k_{B}T_{h}/n_{c,0}e^{2}\right)  ^{1/2}$,
the inverse cool electron plasma frequency $\omega_{pc}^{-1} $ = $(\varepsilon
_{0}m_{e}/$ $n_{c,0}e^{2})^{1/2}$, the potential scale reads $\phi_{0}=k_{B}
T_{h}/e$, and the thermal pressures scale $P_{0}=n_{c,0}k_{B}T_{c}$. We also
define the mass ratio of electron to ion as $\mu=m_{e}/m_{i}=1/1836$ (proton)
and the number of ions as $Z=1$ (Hydrogen).

\section{Linear Method}\label{twofluidmodel:dr}

Let us assume that $S=(n,u,\tilde{n},\tilde{u},P,\phi)$ be the system
variables that describe the system's state at a given space and time. The
small deviations from the equilibrium state are $S^{(0)}=(1,0,1+\beta,0,1,0)$.
We use the first-order derivatives of the harmonic wave amplitude as Eq.
(\ref{eq1_17}), we get the following expressions for velocity, density of the
cool electrons and the ions, and thermal pressure,
\begin{equation}
\begin{array}
[c]{ccc}
n_{1}^{(1)}=\dfrac{k}{\omega}u_{1}^{(1)},\text{  } & u_{1}^{(1)}=-\dfrac
{k}{\omega}\left(  \phi_{1}^{(1)}-\sigma P_{1}^{(1)}\right)  ,\text{  } &
P_{1}^{(1)}=3n_{1}^{(1)},
\end{array}
\label{eq3_21}
\end{equation}
\begin{equation}
\begin{array}
[c]{cc}
\tilde{n}_{1}^{(1)}=\dfrac{k}{\omega}\tilde{u}_{1}^{(1)},\text{  } &
\tilde{u}_{1}^{(1)}=\mu\dfrac{k}{\omega}\phi_{1}^{(1)},
\end{array}
\label{eq3_23}
\end{equation}
The Poisson's equation closes all system variables together.
\begin{equation}
-k^{2}\phi_{1}^{(1)}=-\beta-\tilde{n}_{1}^{(1)}+n_{1}^{(1)}+\beta\left(
1-\frac{\phi}{(\kappa-\tfrac{3}{2})}\right)  ^{-\kappa+1/2}. \label{eq3_24}
\end{equation}
Using the fact that $\mu\ll1$, we use the Taylor expansion to first order. If
we approximate to first order, we obtain the linear dispersion relation
$\omega_{3}=\omega_{3}(k)$:
\begin{equation}
\omega_{3}^{2}(k)\simeq\omega_{2}^{2}(k){+}\frac{{\mu k^{2}}}{\left(
k^{2}+k_{D}^{2}\right)  \left[  1+3{\sigma}\left(  k^{2}+k_{D}^{2}\right)
\right]  }, \label{eq3_27}
\end{equation}
where $k_{D}$ is defined by Eq. (\ref{eq1_22}), and $\omega_{2}(k)$, the wave
frequency of the one-fluid warm model, is given by Eq. (\ref{eq2_21}). In the
limit $\mu\rightarrow0$, we get the one-fluid warm model as Eq. (\ref{eq2_21}).

\subsection{Ion Inertia Effects on Linear Waves}

\begin{figure}
\begin{center}
\includegraphics[width=6.0in]{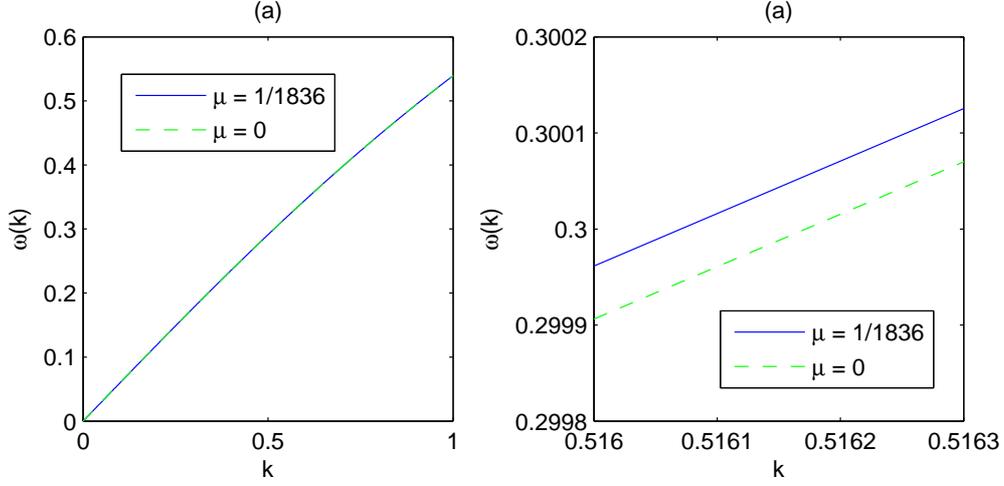}
\caption{Variation of the dispersion function curve for different values of
 $\mu$. Curves from top to bottom: $\mu=1/1836$ (solid), and $0$ (dashed).
Here, $\kappa=3$, $\beta=2$, $\sigma=0.02$, and $Z=1$.}
\label{fig3_3}
\end{center}
\end{figure}

\begin{figure}
\begin{center}
\includegraphics[width=4.0in]{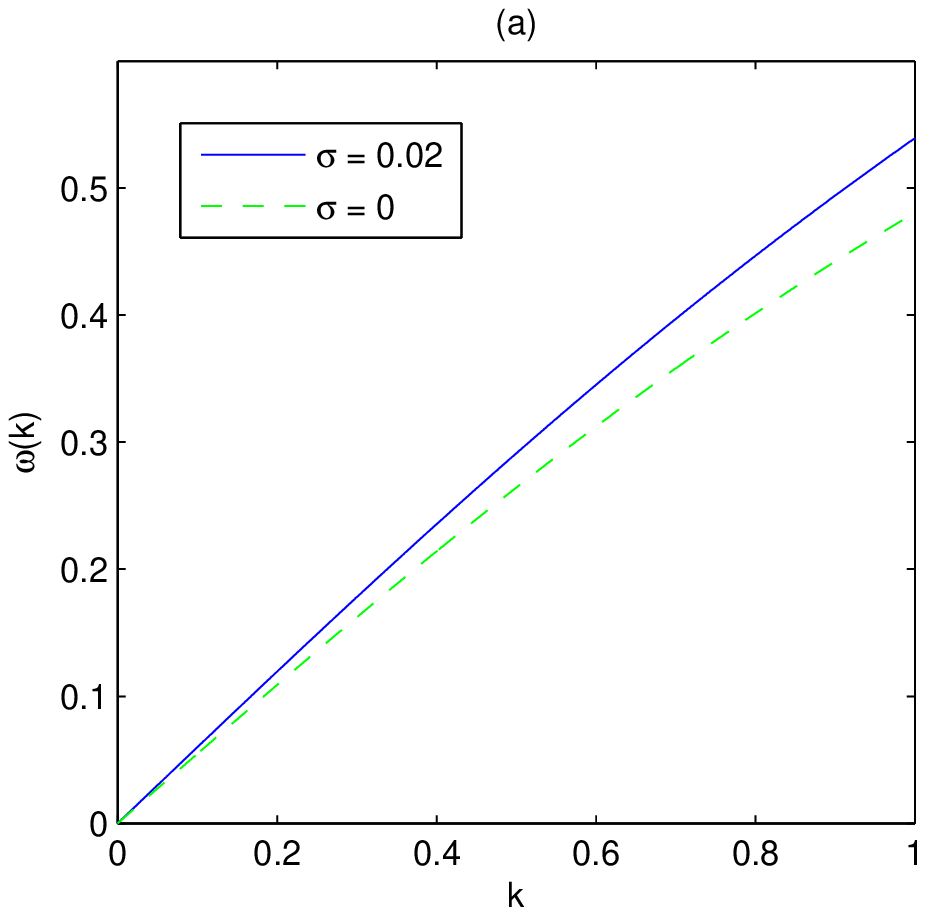}
\caption[ariation of the dispersion function curve for different values of
 $\sigma$.]{Variation of the dispersion function curve for different values of
 $\sigma$. Curves from top to bottom: $\sigma=0.02$ (solid), and
$0$ (dashed). Here, $\kappa=3$, $\beta=2$, $Z=1$, and $\mu=1/1836$.}
\label{fig3_4}
\end{center}
\end{figure}

To understand how inertial ions affect the linear dispersion function, we may
write Eq. (\ref{eq3_27}) as follows
\begin{equation}
\omega_{3}^{2}\simeq\left(  1+\frac{{\mu}}{1+3{\sigma}\left(  k^{2}+k_{D}%
^{2}\right)  }\right)  \omega_{1}^{2}+3{\sigma}k^{2}.
\end{equation}
We see that the thermal effect has an dramatic effect on the results of the
inertial ions. Hence, there is extremely small difference between the
dispersion curve of this model and the model described in \S  \ref{warmmodel}, as shown in Fig. \ref{fig3_3}. In the limit $\sigma\rightarrow0$, we obtain
$\omega_{3}\simeq\left(  1+{\mu}\right)  ^{1/2}\omega_{1}\approx\left(
1+\frac{1}{2}{\mu}\right)  \omega_{1}$, with the result that the
electron-acoustic phase speed increases by order of $\frac{1}{2}{\mu}$ (for
the Hydrogen $\mu=m_{e}/m_{i}=1/1836$). Figure \ref{fig3_4} shows the
difference between two-fluid warm model ($\sigma=0.02$) and two-fluid cold
model ($\sigma=0$). We see that the thermal effect ($\sqrt{3{\sigma}}$) plays
a role in modifying the dispersion curve more than the inertial ions (while
$\mu\ll1$). It seems that the inertial ions make some minor effects to the
electron-acoustic phase speed.

\section{Nonlinear Pseudopotential Technique}\label{twofluidmodel:nonlinear}

To investigate the existence of the electron-acoustic solitary waves, we use
the pseudopotential approach by assuming that all dependent variables depend
on the traveling coordinate $\xi=x-Mt$, where $M$ is the Mach number. Using
this transformation, we get Eqs. (\ref{eq2_27})--(\ref{eq2_28_1}), and the
ion-fluid equations take the following form:
\begin{equation}
-M\dfrac{d\tilde{n}}{d\xi}+\frac{d(\tilde{n}\tilde{u})}{d\xi}=0,
\label{eq3_35}
\end{equation}
\begin{equation}
-M\dfrac{d\tilde{u}}{d\xi}+\tilde{u}\dfrac{d\tilde{u}}{d\xi}=-\mu\frac{d\phi
}{d\xi}, \label{eq3_36}
\end{equation}
\begin{equation}
\frac{d^{2}\phi}{d\xi^{2}}=-\tilde{n}+n+\beta\left(  1-\frac{\phi}%
{(\kappa-\tfrac{3}{2})}\right)  ^{-\kappa+1/2}. \label{eq3_37}
\end{equation}

Integrating Eqs. (\ref{eq2_27})--(\ref{eq2_28_1}) and Eqs. (\ref{eq3_35})--(\ref{eq3_36}) yield
\begin{equation}
\begin{array}
[c]{cc}
u=M(1-\dfrac{1}{n}), & u={M-}\left(  {M}^{2}{+2\phi-3n^{2}\sigma+3\sigma
}\right)  ^{1/2},
\end{array}
\label{eq3_38}
\end{equation}
\begin{equation}
\begin{array}
[c]{cc}
\tilde{u}=M\left(  1-\dfrac{(1+\beta}{\tilde{n}}\right)  , & \tilde{u}
=M{-}\left(  {M}^{2}{-2\mu\phi}\right)  ^{1/2}.
\end{array}
\label{eq3_40}
\end{equation}
Combining Eqs. (\ref{eq3_38})--(\ref{eq3_40}), we get
\begin{equation}
{n=}\dfrac{1}{2\sqrt{3{\sigma}}}\left(  \left[  {2\phi+}\left(  {M+}
\sqrt{3{\sigma}}\right)  ^{2}\right]  ^{1/2}\pm\left[  {2\phi+\left(
{M-}\sqrt{3{\sigma}}\right)  ^{2}}\right]  ^{1/2}\right)  , \label{eq3_41}%
\end{equation}
\begin{equation}
\tilde{n}=(1+\beta)\left(  {1-\mu}\frac{{2\phi}}{{M}^{2}}\right)  ^{-1/2}.
\label{eq3_43}
\end{equation}
The upper/lower sign in Eq. (\ref{eq3_41}) is for subsonic/supersonic
solitons, respectively. In the limit $\mu\rightarrow0$, we recover the
inertialess ions ($\tilde{n}=1+\beta$). We also obtain the condition at
equilibrium ($n=1$ and $\tilde{n}=1+\beta$) through the limit $\phi
\rightarrow0$.

Eq. (\ref{eq3_41}) shows that $\phi_{\max}^{(-)}=-\frac{1}{2}\left(  {M-}
\sqrt{3{\sigma}}\right)  ^{2}$, which is considered to be the maximum (in
absolute value) limit for the negative electrostatic wave potential.
Meanwhile, two-fluid model, Eq. (\ref{eq3_43}), gives a maximum limit for the
positive electrostatic wave potential $\phi_{\max}^{(+)}=\frac{1}{{2}}{\mu
}^{-1}{{M}^{2}}$. We can see that the maximum limit for the positive solitary
waves is in proportion to ${\mu}^{-1}$ (for the proton $\mu^{-1}=1836$). This
means that the two-fluid model may support a positive soliton with very large
amplitude (by order of ${\mu}^{-1}$) in comparison with negative solitons.
However, we must also think of the possible range of the propagation velocity
($M$), which is valid for the positive solitary waves. In the two-fluid model,
the positive pulses usually appear to be subsonic ($M<1$), i.e., heavy species
(ion) propagating slowly. Hence, we may not observe very large positive pulses
due to small velocity ($M\ll1$).

Substituting equations (\ref{eq3_41}) and (\ref{eq3_43}) into equation
(\ref{eq3_37}), we get the equation of motion:
\begin{align}
\frac{d^{2}\phi}{d\xi^{2}}  &  =-\Psi_{3}^{\prime}(\phi,M,\beta,\kappa
,{\sigma},\mu)=-(1+\beta)\left(  {1-\mu}\frac{{2\phi}}{{M}^{2}}\right)
^{-1/2}+\beta\left(  1-\frac{\phi}{(\kappa-\tfrac{3}{2})}\right)
^{-\kappa+1/2}\nonumber\\
&  +\dfrac{1}{2\sqrt{3{\sigma}}}\left(  \left[  {2\phi+}\left(  {M+}
\sqrt{3{\sigma}}\right)  ^{2}\right]  ^{1/2}\pm\left[  {2\phi+\left(
{M-}\sqrt{3{\sigma}}\right)  ^{2}}\right]  ^{1/2}\right)  . \label{eq3_47}
\end{align}
Multiplying the above equation by $d\phi/d\xi$, integrating, and applying
boundary condition, namely $n=1$, $\tilde{n}=1+\beta$, $P=1$, and $u=\tilde
{u}=\phi=0$, we find the energy balance equation:
\begin{equation}
\frac{1}{2}\left(  \frac{d\phi}{d\xi}\right)  ^{2}+\Psi_{3}(\phi
,M,\beta,\kappa,{\sigma},\mu)=0, \label{eq3_48}
\end{equation}
where the Sagdeev pseudopotential $\Psi_{3}(\phi,M,\beta,\kappa,{\sigma},\mu)$
is written as
\begin{align}
\Psi_{3}(\phi,M,\beta,\kappa,{\sigma},\mu)  &  =(1+\beta)\frac{{M}^{2}}{{\mu}
}\left(  1-\left(  {1-\mu}\frac{{2\phi}}{{M}^{2}}\right)  ^{1/2}\right)\nonumber\\
&+\beta\left(  1-\left(  1-\frac{\phi}{\kappa-\tfrac{3}{2}}\right)
^{-\kappa+3/2}\right) \nonumber\\
&  +\frac{1}{6\sqrt{3{\sigma}}}\left(  \left(  {M+}\sqrt{3{\sigma}}\right)
^{3}\pm{{\left(  {M-}\sqrt{3{\sigma}}\right)  ^{3}}}\right. \nonumber\\
&  \left.  -\left[  {2\phi+}\left(  {M+}\sqrt{3{\sigma}}\right)  ^{2}\right]
^{3/2}\mp{\left[  {2\phi+\left(  {M-}\sqrt{3{\sigma}}\right)  ^{2}}\right]
}^{3/2}\right)  . \label{eq3_49}
\end{align}
In the limit $\mu\rightarrow0$, we obtain one-fluid warm model, i.e.,
$\lim_{\mu\rightarrow0}\Psi_{3}(\phi,M,\beta,\kappa,{\sigma},\mu)=\Psi
_{2}(\phi,M,\beta,\kappa,{\sigma})$ as in Eq. (\ref{eq2_38}). We also get the
cold model from the limit ${\sigma}\rightarrow0$ (see Eq. (\ref{eq1_21}))

\subsection{Ion Inertia Effects on EA Solitons}

\begin{figure}
\begin{center}
\includegraphics[width=5.0in]{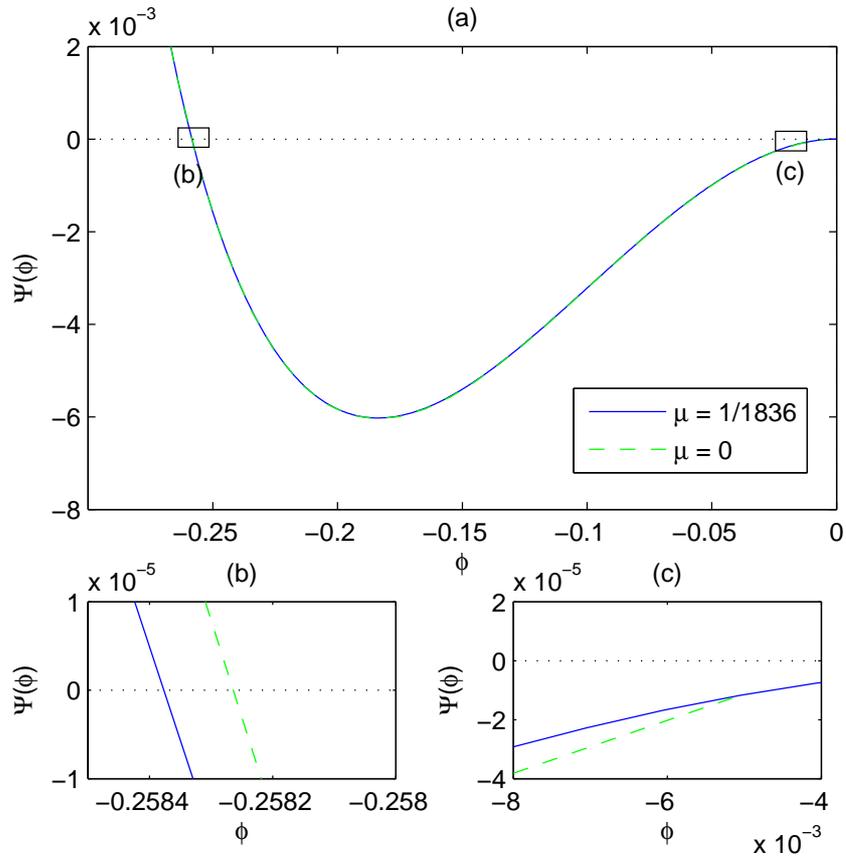}
\caption[Variation of pseudopotential $\Psi(\phi)$ with $\phi$ for
$\mu=1/1836$ and $0$.]{(a) Variation of pseudopotential $\Psi(\phi)$ with $\phi$ for
$\mu=1/1836$ (solid) and $0$ (dot-dashed curve). As zoomed in on (b) and (c),
difference between two curves are extremely small due to small value of $\mu$.
Here, $\beta=1.1$, $\kappa=3$, $M=1$, and $Z=1$.}
\label{fig3_13}
\end{center}
\end{figure}

\begin{figure}
\begin{center}
\includegraphics[width=5.0in]{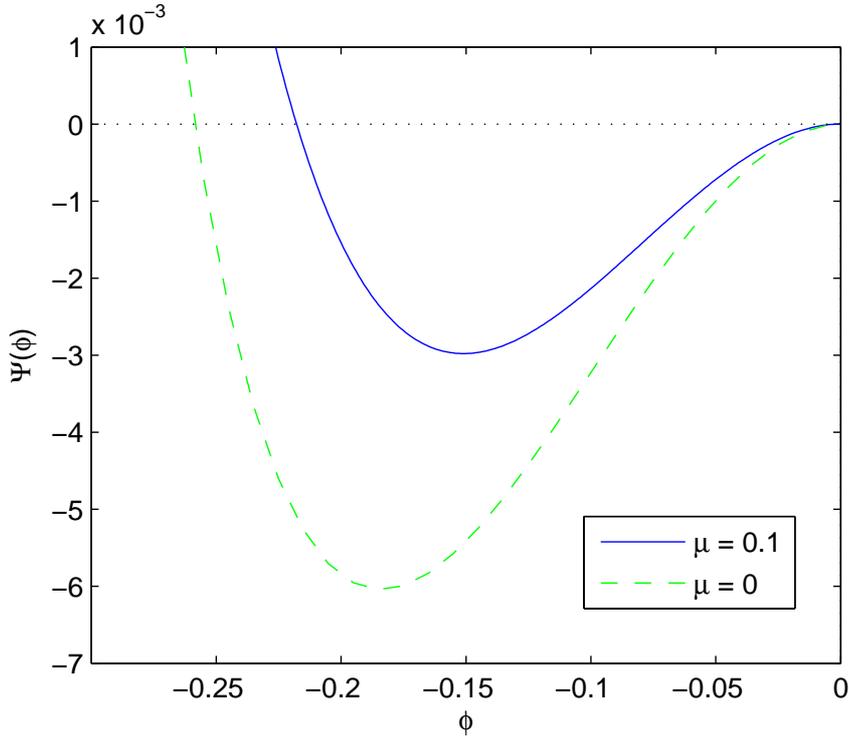}
\caption[Variation of pseudopotential $\Psi(\phi)$ with $\phi$ for different
mass ratio $\mu$.]{Variation of pseudopotential $\Psi(\phi)$ with $\phi$ for different
mass ratio $\mu$. Curves from top to bottom: $\mu=0.1$ (solid) and $0$
(dot-dashed curve). Here, $\mu=0.1$ has not physical mean and other parameters
are same as used in Fig. \ref{fig3_13}.}
\label{fig3_14}
\end{center}
\end{figure}

As illustrated in Fig. \ref{fig3_13}, the ion-fluid has a trivial role in
modifying negative supersonic (${M>}\sqrt{3{\sigma}}$) solitary waves. Figure
\ref{fig3_14} shows the difference between two-fluid model for $\mu=0.1$ and
one-fluid model. However, $\mu=0.1$ has not physical mean, and only was used
to distinguish between them.

\begin{figure}
\begin{center}
\includegraphics[width=6.0in]{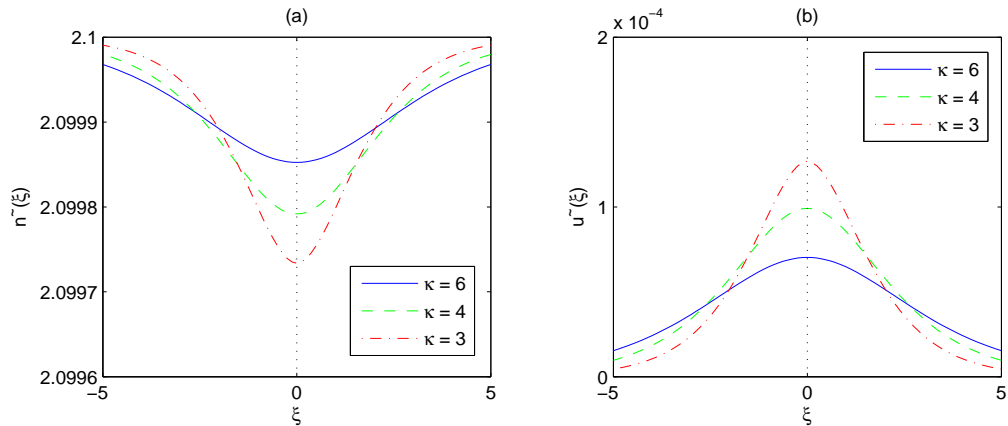}
\caption[(a) Variation of density $\tilde{n}$ with $\xi$ for different
$\kappa$. (b) Variation of velocity $\tilde{u}$ with $\xi$ for different
$\kappa$.]{(a) Variation of density $\tilde{n}$ with $\xi$ for different
$\kappa$. (b) Variation of velocity $\tilde{u}$ with $\xi$ for different
$\kappa$. Curves from top to bottom: $\kappa=6$ (solid), $4$ (dashed),
$3$ (dot-dashed curve). Here, $\sigma=0.02$, $\beta=1.1$, $M=1$, $Z=1$, and
$\mu=1/1836$.}
\label{fig3_11}
\end{center}
\end{figure}

\begin{figure}
\begin{center}
\includegraphics[width=6.0in]{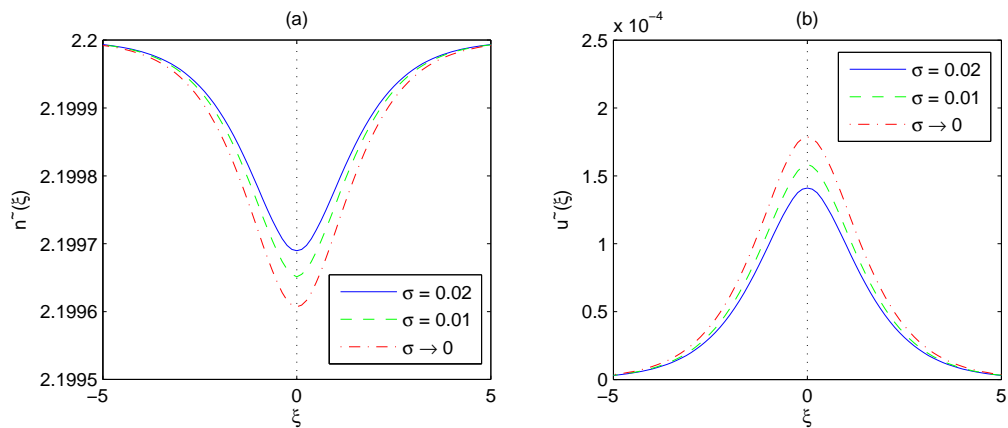}
\caption[(a) Variation of density $\tilde{n}$ with $\xi$ for different
temperature ratio $\sigma$. (b) Variation of velocity $\tilde{u}$ with $\xi
$ for different temperature ratio $\sigma$.]{(a) Variation of density $\tilde{n}$ with $\xi$ for different
temperature ratio $\sigma$. (b) Variation of velocity $\tilde{u}$ with $\xi
$ for different temperature ratio $\sigma$. Curves from top to bottom:
$\sigma\rightarrow0$ (dot-dashed curve), $\sigma=0.01$ (dashed),
$0.02$ (solid). Here, $\beta=1.1$, $\kappa=3$, $M=1$, $Z=1$, and $\mu
=1/1836$.}
\label{fig3_7}
\end{center}
\end{figure}

Numerically solving Eq. (\ref{eq3_49}) provides the number density and the
velocity of the ions. Figure \ref{fig3_7} shows the variation of $\tilde{n}$
and $\tilde{u}$ for different temperature ratio $\sigma$ are slight. We see a
decline in the absolute ion quantities (density and velocity) with an increase
in the thermal velocity $\sqrt{3{\sigma}}$. Figure \ref{fig3_11} shows the
variation of the ion density and the ion velocity for different $\kappa$. We
note that, by increasing $\kappa$ (closer to the Maxwellian background), the
ion quantities decreases. Hence, the inertial ions are more affect by
suprathermal species than the Maxwellian distribution

\subsection{Positive Solitary Wave Structure}

\begin{figure}
\begin{center}
\includegraphics[width=6.0in]{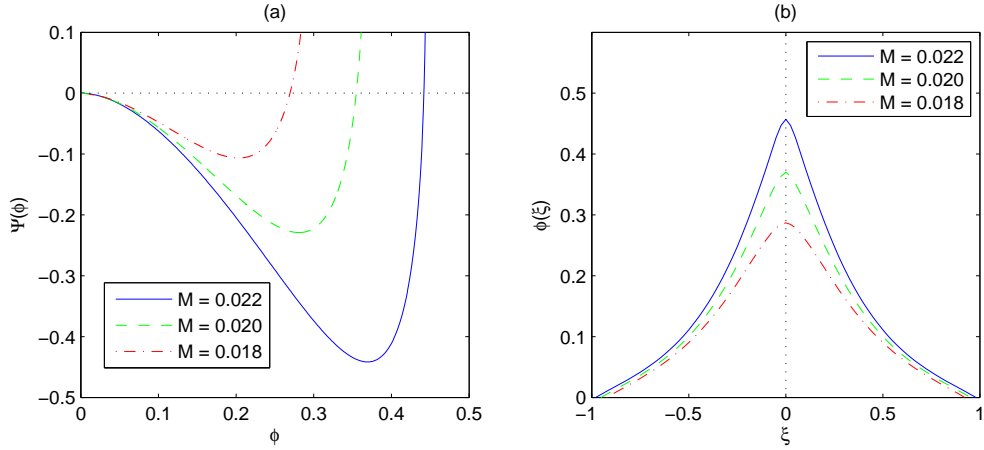}
\caption[Compressive (positive) solitary structures at subsonic region
($M\ll\Theta$).]{Compressive (positive) solitary structures at subsonic region
($M\ll\Theta$). (a) Variation of pseudopotential $\Psi(\phi)$ with $\phi
$ for different values of the Mach number $M$. (b) Variation of potential
$\phi$ with $\xi$ for different $M$. Curves from top to bottom:
$M=0.022$ (solid), $0.020$ (dashed), $0.018$ (dot-dashed curve). Here,
$\beta=3$, $\kappa=3$, $\sigma=0.02$, $Z=1$, and $\mu=1/1836$.}
\label{fig3_15}
\end{center}
\end{figure}

\begin{figure}
\begin{center}
\includegraphics[width=6.0in]{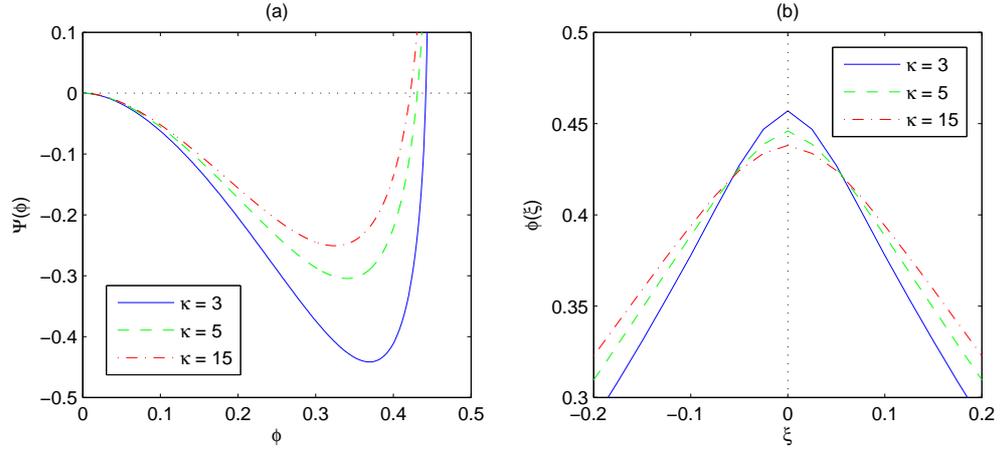}
\caption[Compressive solitary structures at subsonic region.]{Compressive solitary structures at subsonic region. (a) Variation of
pseudopotential $\Psi(\phi)$ with $\phi$ for different $\kappa$. (b)
Variation of potential $\phi$ with $\xi$ for different $\kappa$. Curves from
top to bottom: $\kappa=3$ (solid), $5$ (dashed), $15$ (dot-dashed curve).
Here, $M=0.022$, $\beta=3$, $\kappa=3$, $\sigma=0.02$, $Z=1$, and $\mu
=1/1836$.}
\label{fig3_16}
\end{center}
\end{figure}

It is interesting to see the subsonic solution ($M<\sqrt{3{\sigma}}$), which
is associated with the upper sign in Eq. (\ref{eq3_49}). Previously
(\S  \ref{warmmodel:nonlinear}), we classified the Mach number under two
regions, i.e., subsonic/supersonic for hot/cool species, respectively. The
cool electron-fluid can generally support a negative supersonic electrostatic
wave. But, ion-fluid may possess a subsonic soliton, which gives a positive
pulse. We have numerically solved Eq. (\ref{eq3_49}) for the subsonic
condition. As illustrated in Fig. \ref{fig3_15}, this makes the positive
electrostatic wave potential. We see that the amplitude of pulse rises as the
Mach number is increased. Figure \ref{fig3_16} shows that increasing $\kappa$
(approach the Maxwellian distribution) reduces the positive solitary pulse
amplitude, but extends the full width at half maximum (FWHM).

\begin{figure}
\begin{center}
\includegraphics[width=5.0in]{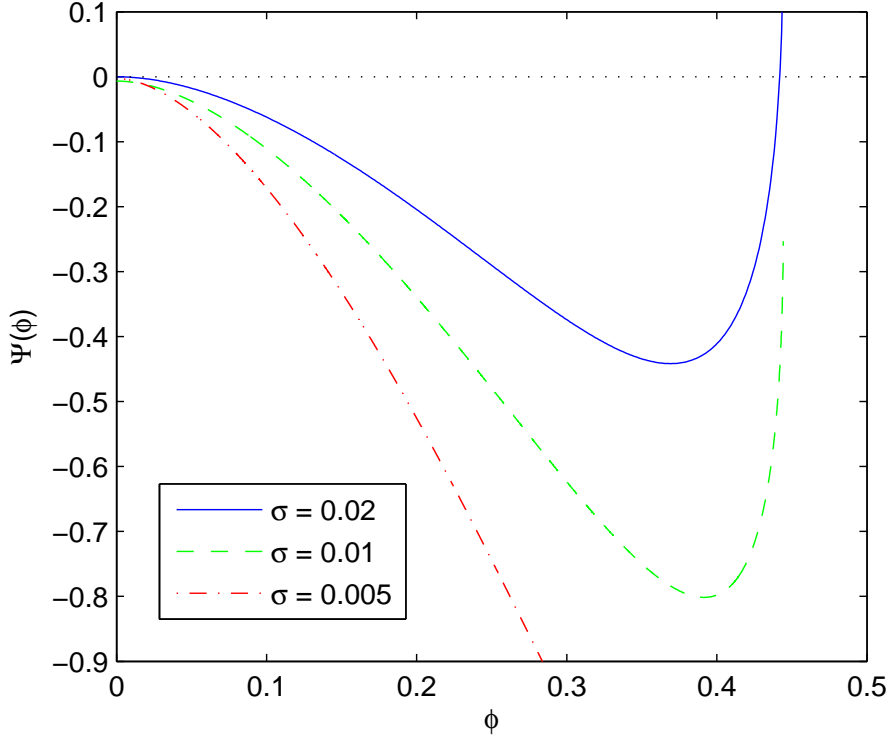}
\caption[Variation of pseudopotential $\Psi(\phi)$ with $\phi$ for
different temperature ratio $\sigma$.]{Variation of pseudopotential $\Psi(\phi)$ with $\phi$ for
different temperature ratio $\sigma$. Curves from bottom to top: $\sigma
=0.02$ (solid), $\sigma=0.01$ (dashed), $0.005$ (dot-dashed curve). Here,
$M=0.022$, $\beta=3$, $\kappa=3$, $Z=1$, and $\mu=1/1836$.}
\label{fig3_17}
\end{center}
\end{figure}

\begin{figure}
\begin{center}
\includegraphics[width=4.5in]{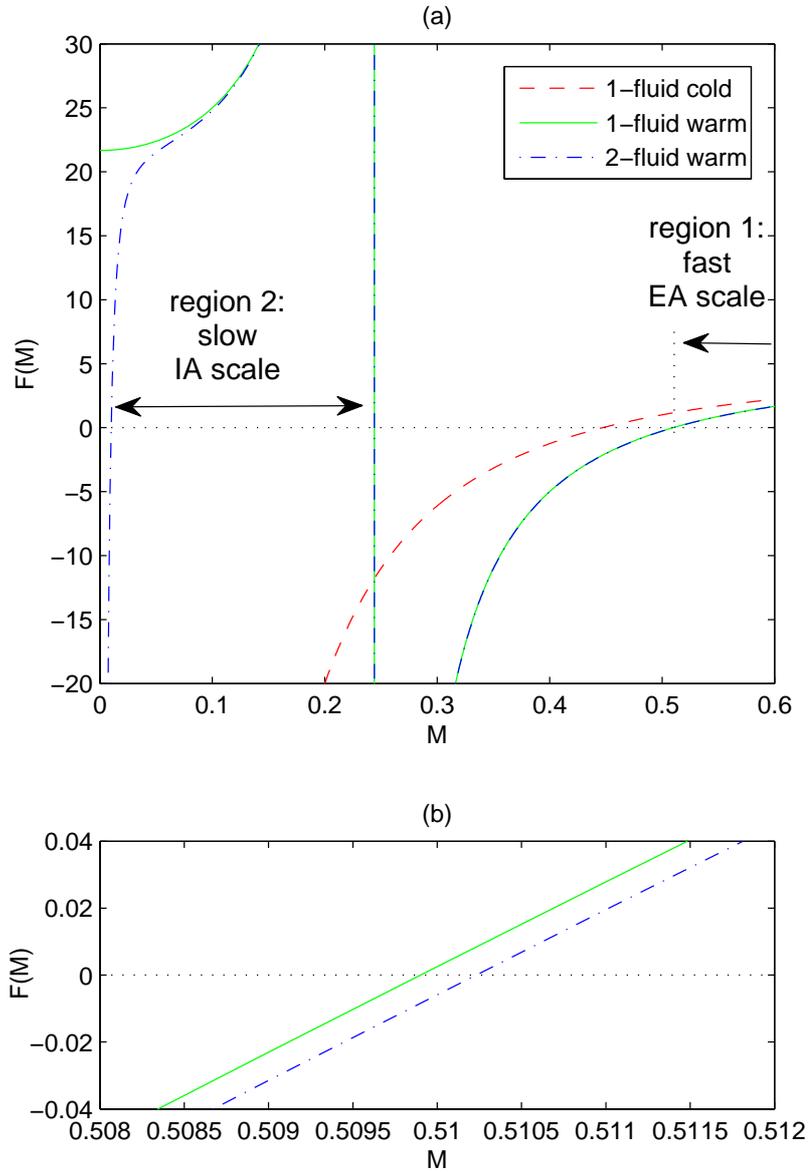}
\caption[The existence domains for stationary solitary structures.]{The existence domains for stationary solitary structures. The
quantities $F_{1}$ for 1-fluid cold model  (dashed curve), 1-fuild warm model
(solid), and 2-fluid warm model (dot-dashed) are defined in (\ref{eq1_53}),
(\ref{eq2_44}) and (\ref{eq3_56}), respectively. As shown in (a), the 2-fluid
warm model has two existence domains, namely the (fast) electron-acoustic (EA)
scale, and the (slow) ion-acoustic (IA) scale. As zoomed in on (b), difference
between 1-fuild warm model and 2-fluid warm model are extremely small in
supersonic region ($M>\sqrt{3\sigma}$) due to small value of $\mu$. Here,
$\kappa=3$, $\beta=3$, $\sigma=0.02$, $Z=1$, and $\mu=1/1836$. }
\label{fig3_21}
\end{center}
\end{figure}

It is important to note that the two-fluid cold model ($\sigma\rightarrow0$)
may not produce the positive solitary structures. In the limit $\sigma
\rightarrow0$, the number density (\ref{eq3_41}) approaches $n=\left(
1+2\phi/M^{2}\right)  ^{-1/2}$ and the Sagdeev pseudopotential reads as
\begin{align}
\Psi_{3}(\phi,M,\beta,\kappa,\mu)  &  =(1+\beta)\frac{{M}^{2}}{{\mu}}\left(
1-\left(  {1-\mu}\frac{{2\phi}}{{M}^{2}}\right)  ^{1/2}\right) \nonumber\\
&  +\beta\left(
1-\left(  1-\frac{\phi}{\kappa-\tfrac{3}{2}}\right)  ^{-\kappa+3/2}\right)
\nonumber\\
&  +M^{2}\left(  1-\left(  1+\frac{2\phi}{M^{2}}\right)  ^{1/2}\right)  .
\label{eq3_50}
\end{align}
It is difficult to find a positive solution to Eq. (\ref{eq3_50}) in the same
way as given in Eq. (\ref{eq3_49}). As shown in Fig. \ref{fig3_17}, reducing
the electron thermal velocity affects the positive soliton existence. Indeed,
it seems there is no possibility of positive solitary structure for very small
$\sigma$.

\section{Negative Electron-Acoustic Soliton Existence}\label{twofluidmodel:nonlinear:rarefactiveexistence}

For the existence of negative potential solitons moving at velocity $M$, we
require $\Psi_{3}^{\prime}(\phi,M,\beta,\kappa,{\sigma},\mu)|_{\phi=0}=0$ and
$\Psi_{2}^{\prime\prime}(\phi,M,\beta,\kappa,{\sigma})|_{\phi=0}<0$. Hence,
the lower Mach number limit can be obtained through the following function:
\begin{equation}
F_{1}(M,\beta,\kappa,{\sigma},\mu)=-\Psi_{3}^{\prime\prime}
\vert_{\phi=0}=\frac{\beta
(\kappa-\frac{1}{2})}{\kappa-\tfrac{3}{2}}-\frac{1}{M^{2}-3\sigma}
-(1+\beta)\frac{{\mu}}{{M}^{2}}>0. \label{eq3_56}
\end{equation}
Eq. (\ref{eq3_56}) leads to graphs where the existence domains for stationary
solitary structures are illustrated. As shown in Fig. \ref{fig3_21}, the
thermal velocity classifies the Mach number under two regions, namely
\textquotedblleft fast\textquotedblright (${M>}\sqrt{{{3}}\sigma}$) and
\textquotedblleft slow\textquotedblright (${M<}\sqrt{{{3}}\sigma}$) scales,
i.e., the thermal velocity is smaller or larger than the Mach number,
respectively. We will see that the thermal velocity divides the propagation
speed into two ranges: negative and positive solitary waves.

Eq. (\ref{eq3_56}) provides the lower Mach number limit for negative solitary
structures (note: Eq. (\ref{eq3_56}) gives us two solutions; also Eq.
(\ref{eq3_60})):
\begin{align}
M_{1}^{(-)}(\beta,\kappa,{\sigma},\mu)  &  =\frac{1}{2}\left(  \frac
{(\kappa-\tfrac{3}{2})\left[  {\mu}(1+\beta)+1\right]  }{\beta(\kappa-\frac
{1}{2})}+3\sigma+2\left(  \frac{3\sigma{\mu}(1+\beta)(\kappa-\tfrac{3}{2}
)}{\beta(\kappa-\frac{1}{2})}\right)  ^{1/2}\right)  ^{1/2}\nonumber\\
&  +\frac{1}{2}\left(  \frac{(\kappa-\tfrac{3}{2})\left[  {\mu}(1+\beta
)+1\right]  }{\beta(\kappa-\frac{1}{2})}+3\sigma-2\left(  \frac{3\sigma{\mu
}(1+\beta)(\kappa-\tfrac{3}{2})}{\beta(\kappa-\frac{1}{2})}\right)
^{1/2}\right)  ^{1/2}. \label{eq3_58}
\end{align}
In the limit $\mu\rightarrow0$, we get the same expression (\ref{eq2_45}) for
the one-fluid warm model.

\begin{figure}
\begin{center}
\includegraphics[width=6.0in]{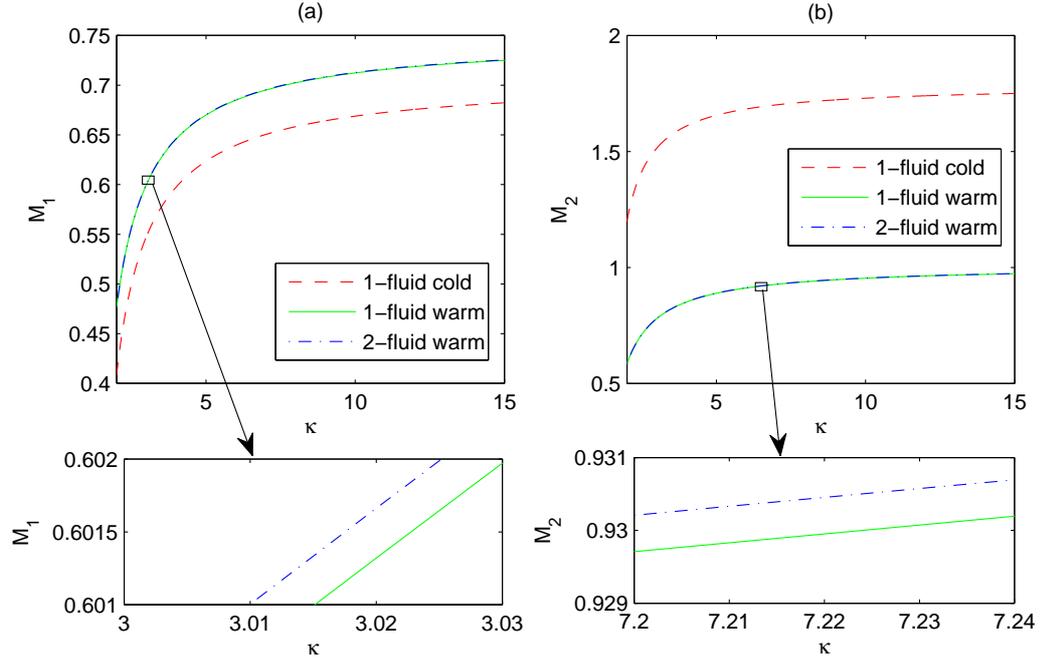}
\caption[Variation of the lower limit $M_{1}$ and the upper limit $M_{2}$ with $\kappa$ for 1-fluid cold model, 1-fluid warm model, and 2-fluid warm model.]{Variation of the lower limit $M_{1}$ and the upper limit $M_{2}$ with $\kappa$ for 1-fluid cold model (dashed curve), 1-fluid warm model
(solid), and 2-fluid warm model (dot-dashed). As zoom-in shows difference between
1-fluid warm model and 2-fluid warm model are extremely small. Here, $\beta
=2$, $\sigma=0.02$, $Z=1$, and $\mu=1/1836$. }
\label{fig3_20}
\end{center}
\end{figure}

We obtain the higher limit for the Mach number through $F_{2}(M,\beta
,\kappa,{\sigma},\mu)=\Psi_{3}(\phi,M,\beta,\kappa,{\sigma},\mu)|_{\phi
=\phi_{\max}^{(-)}}>0$, where $\phi_{\max}^{(-)}=-\frac{1}{2}\left(  {M-}\sqrt{3{\sigma}}\right)  ^{2}$. This gives:
\begin{align}
F_{2}^{(-)}(M,\beta,\kappa,{\sigma},\mu)  &  =(1+\beta)\frac{{M}^{2}}{{\mu}
}\left(  1-\left(  {1+\mu}+\frac{{\mu(3{\sigma}-2M}\sqrt{3{\sigma}})}{{M}^{2}
}\right)  ^{1/2}\right) \nonumber\\
&  +\beta\left(  1-\left(  1+\frac{\left(  {M-}\sqrt{3{\sigma}}\right)  ^{2}
}{2\kappa-3}\right)  ^{-\kappa+3/2}\right)  +M^{2}+\sigma-\frac{4M\sqrt
{M\sqrt{3{\sigma}}}}{3}. \label{eq3_59}
\end{align}
Hence, the upper limit $M_{2}^{(-)}(\beta,\kappa,{\sigma},\mu)$ is obtained by
solving the above equation.

\subsection{Ion Inertia Effects on Negative Soliton}

Figure \ref{fig3_20} shows that the ion inertia effects have trivially
negative soliton existence altered. We notice that there is a extremely small
difference between one-fluid warm model and two-fluid warm model. At the
supersonic domain, positively charged heavy species behave like uniformly
distributed positive background with negligible role in the dynamics of EAWs.

\section{Positive Electron-Acoustic Soliton Existence}\label{twofluidmodel:nonlinear:compressiveexistence}

However, Eq. (\ref{eq3_56}) has another solution, which yields the lower Mach
number limit for positive solitary structures:
\begin{align}
M_{1}^{(+)}(\beta,\kappa,{\sigma},\mu)  &  =\frac{1}{2}\left(  \frac
{(\kappa-\tfrac{3}{2})\left[  {\mu}(1+\beta)+1\right]  }{\beta(\kappa-\frac
{1}{2})}+3\sigma+2\left(  \frac{3\sigma{\mu}(1+\beta)(\kappa-\tfrac{3}{2}
)}{\beta(\kappa-\frac{1}{2})}\right)  ^{1/2}\right)  ^{1/2}\nonumber\\
&  -\frac{1}{2}\left(  \frac{(\kappa-\tfrac{3}{2})\left[  {\mu}(1+\beta
)+1\right]  }{\beta(\kappa-\frac{1}{2})}+3\sigma-2\left(  \frac{3\sigma{\mu
}(1+\beta)(\kappa-\tfrac{3}{2})}{\beta(\kappa-\frac{1}{2})}\right)
^{1/2}\right)  ^{1/2}. \label{eq3_60}
\end{align}
It is interesting to see that $\lim_{\mu\rightarrow0}M_{1}^{(+)}(\beta
,\kappa,{\sigma},\mu)=0$. This means that the one-fluid model involving
inertial (cold or cool) electrons and inertialess ions may not produce
positive solitons due to the dynamics of positively charged species being
negligible.

We also derive the upper Mach number limit from $F_{2}(M,\beta,\kappa,{\sigma
},\mu)$ = $\Psi_{3}(\phi,$ $M,\beta,\kappa,{\sigma},\mu)|_{\phi=\phi_{\max}^{(+)}}
>0$, where $\phi_{\max}^{(+)}=\frac{1}{{2}}{\mu}^{-1}{{M}^{2}}$. This yields
the following equation:
\begin{align}
F_{2}^{(+)}(M,\beta,\kappa,{\sigma},\mu)  &  =(1+\beta)\frac{{M}^{2}}{{\mu}
}+\beta\left(  1-\left(  1-\frac{{{M}^{2}}}{{2\mu}\left(  \kappa-\tfrac{3}
{2}\right)  }\right)  ^{-\kappa+3/2}\right) \nonumber\\
&  +\frac{1}{6\sqrt{3{\sigma}}}\left(  \left(  {M+}\sqrt{3{\sigma}}\right)
^{3}+{{\left(  {M-}\sqrt{3{\sigma}}\right)  ^{3}}}\right. \nonumber\\
&  \left.  -\left[  {\frac{1}{{\mu}}{{M}^{2}}+}\left(  {M+}\sqrt{3{\sigma}
}\right)  ^{2}\right]  ^{3/2}-{\left[  {\frac{1}{{\mu}}{{M}^{2}}+\left(
{M-}\sqrt{3{\sigma}}\right)  ^{2}}\right]  }^{3/2}\right)  . \label{eq3_61}
\end{align}
In the limit $\mu\rightarrow0$, we find no solution to Eq. (\ref{eq3_61}).
This confirms our previous statement that the one-fluid model described in
\S  \ref{singlecold} and \S  \ref{warmmodel} cannot produce positive
solitary waves.

\subsection{Hot Electron Effects on Positive Soliton}

\begin{figure}
\begin{center}
\includegraphics[width=6.0in]{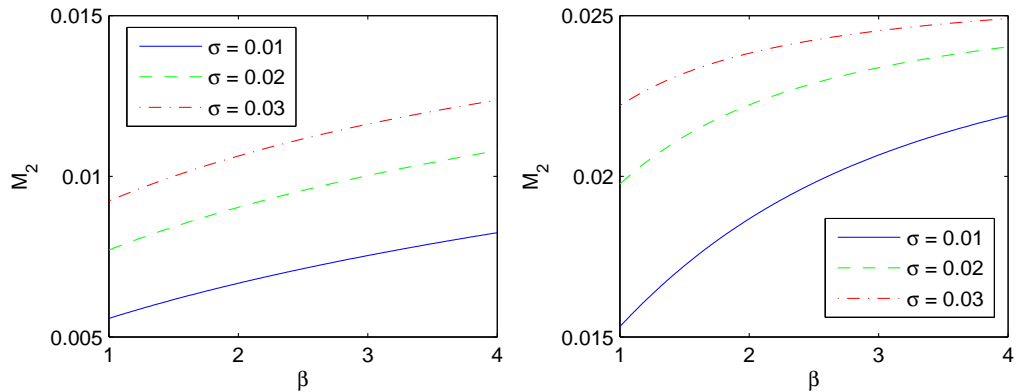}
\caption[Positive potential soliton existence domain in the parameter space of
$\beta$ and Mach number $M$ for different temperature ratio $\sigma$.]{Positive potential soliton existence domain in the parameter space of
$\beta$ and Mach number $M$ for different temperature ratio $\sigma$. (a)
Variation of the lower limit $M_{1}$, (b) Variation of the upper limit $M_{2}$. Curves from bottom to top: $\sigma=0.01$ (solid), $\sigma=0.02$ (dashed), $0.03$ (dot-dashed curve). Here, $\kappa=3$, $Z=1$, and
$\mu=1/1836$.}
\label{fig3_19}
\end{center}
\end{figure}

Fig. \ref{fig3_19} shows that $M_{1}$ and $M_{2}$ rise with the increase in
the parameter $\beta$, i.e., the density of the hot electrons. This result is
in contrast to the negative potential solitary wave (see Fig. \ref{fig2_11}).
The existence domain for the positive potential solitary widens, as the hot
electrons density is increased.

\begin{figure}
\begin{center}
\includegraphics[width=6.0in]{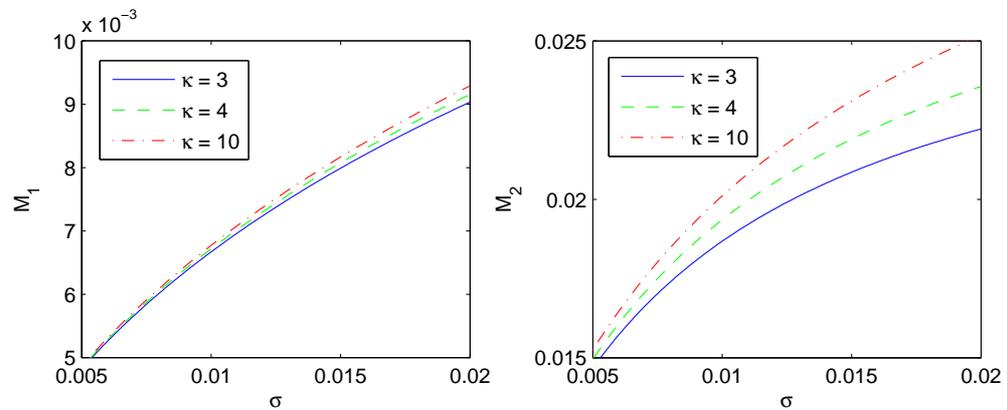}
\caption[Positive potential soliton existence domain in the parameter space of
$\sigma$ and Mach number $M$.]{Positive potential soliton existence domain in the parameter space of
$\sigma$ and Mach number $M$. Solitons may be supported in the region between
$M_{1}(\sigma)$ and $M_{2}(\sigma)$. (a) Variation of $M_{1}(\sigma)$, (b)
Variation of $M_{2}(\sigma)$. Curves from bottom to top: $\kappa=3$ (solid),
$4$ (dashed), $10$ (dot-dashed curve). Here, Here, $\beta=2$, $Z=1$, and
$\mu=1/1836$.}
\label{fig3_18}
\end{center}
\end{figure}

\subsection{Temperature Effects on Positive Soliton}

We also see that the two-fluid cold model ($\sigma\rightarrow0$) may not
propagate the positive solitary pulse, since $\lim_{\sigma\rightarrow0}
M_{1}^{(+)}(\beta,\kappa,{\sigma},\mu)=0$. Numerically solving Eq.
(\ref{eq3_61}) shows that the upper Mach number limit approaches zero in the
limit $\sigma\rightarrow0$. As illustrated in Fig. \ref{fig3_18}, the
existence domain becomes narrower as the thermal velocity is decreased. This
make difficult to find positive solitons at very low $\sigma$. In this figure,
we see that moving into the Maxwellian distribution ($\kappa\rightarrow\infty
$) will increase $M_{1}$ and $M_{2}$.


%
%
%


\chapter{Conclusions}\label{conclusion}

In this research, we have investigated linear and nonlinear EAWs in a
suprathermal plasma consisting of cool (or cold) electrons, in the presence of
hot suprathermal electrons and mobile (or motionless) ions. We began with the
one-fluid cold ($T_{c}=0$) model, and advanced toward the one-fluid warm
($T_{c}\neq0$) in the next step. Including mobile ions, we then approached the
two-fluid warm model. Using small deviations from the equilibrium state to
first order, we have obtained the linear dispersion relation for all three
models. We use a Sagdeev pseudopotential method to investigate nonlinear
structures of the electrostatic solitary waves. Our linear analysis has shown
the weakly damped region for the EAWs, where waves can propagate, under the
influence of hot suprathermal electron, thermal pressure, and ion inertia
effects. Using nonlinear method, we determine the propagation speed and the
existence of stationary profile solitary waves.

In the linear analysis, we found out that growing the suprathermal
distribution, the hot electron number density and temperature, i.e.,
decreasing $\kappa$, increasing $\beta=n_{h,0}/n_{c,0}$, and decreasing
$\sigma=T_{c}/T_{h}$, stretch the weakly damped region. We saw that the
temperature effects dramatically change the dispersion relation. But, the ion
inertia effect is trivial.

We can see that the absolute maximum electrostatic potential increases with
the rise in the suprathermal distribution (decreasing $\kappa$), the hot
electron number density, the hot electron temperature (decreasing $\sigma$).
Nonetheless, the mobile ions have no essential role in the dynamics of
supersonic negative solitary waves. The thermal velocity classifies the Mach
number under two regions, namely supersonic ($M>\sqrt{3\sigma}$) and subsonic
($M<\sqrt{3\sigma}$) ranges. It is interesting to see that the ion-fluid
supports positive subsonic acoustic-solitary waves, while the cool
electron-fluid provides the negative supersonic solitons.

Finally, the nonlinear pseudopotential technique permits existence ranges for
acoustic-solitary waves. The existence domain for the negative potential
soliton becomes narrower with the increase in the suprathermal distribution,
the hot electron number density, and the temperature ratio $\sigma$. The
ion-fluid does not affect the negative soliton existence, but is necessary to
maintain the positive solitary wave structure. The results showed that the
positive acoustic-waves deeply depends on the suprathermal hot electron
parameters ($\kappa$, density, and temperature). We saw that the two-fluid
cold model ($T_{c}=0$) cannot predict the positive solitary pulses. The
existence domain for the positive potential solitary becomes wider, as the hot
electron number density and the temperature ratio $\sigma$ are increased, in
contrast to the results for the negative solitary pulse.

To summarize, Chapter \ref{singlecold} showed how the hot suprathermal
electrons have an effect on the weakly damped region and the propagation
velocity range of the EAWs. The bipolar electric field structures rise as the
hot electron number density is increased. Nonetheless, increasing the hot
electron number density narrows the propagation velocity range. In Chapter
\ref{warmmodel}, we saw how growing the cool temperature increases the damped
region, and decreases the bipolar electric field amplitudes and the soliton
existence. In Chapter \ref{twofluidmodel}, we studied the ion inertia effects
on the EAWs, which does not affect much the negative solitary structures, but
providing positive solitons on the slow scale. We also notice the positive
acoustic solitary waves cannot be propagated in the two-fluid cold model.

To conclude, the electron temperature affects both negative and positive
solitary wave structures. It was found that the mobile ion component has a
trivial role in the supersonic (fast) region, but it appear to be very
important in the subsonic (slow) region, leading to a novel acoustic wave. In
the linear methodology, the ion inertia effect is also negligible. The ion
temperature can be fully included to investigate any different possibilities
(see Appendix \ref{hotionseffect}), but it is beyond the scope of this work.


\bibliographystyle{unsrt}
\addcontentsline{toc}{chapter}{\bibname}




\appendix

\chapter{Analytical Basis}\label{workplan}

\begin{landscape}

The project is composed of three steps (Models 1, 2, and 3):%
\begin{equation}%
\begin{array}
[c]{ccc}%
\text{{\small Model 1 (Chapter \ref{singlecold}):}} & \left\{
\begin{array}
[c]{c}%
\tfrac{\partial n}{\partial t}+\tfrac{\partial(nu)}{\partial x}=0,\\
\\
\tfrac{\partial u}{\partial t}+\tfrac{u\partial u}{\partial x}=\tfrac
{\partial\phi}{\partial x}\underset{\text{{\small Chapter \ref{warmmodel}; Eq.
(\ref{eq2_9})} }}{\underbrace{-\tfrac{\sigma}{n}\tfrac{\partial P}{\partial
x}}},
\end{array}
\right.  &
\begin{array}
[c]{c}%
\text{{\small cool inertial electron continuity }}\\
\text{{\small and momentum equations; }}\\
\text{{\small Eqs. (\ref{eq1_10}) and (\ref{eq1_11})}}%
\end{array}
\\
&  & \\
\text{{\small Model 2 (Chapter \ref{warmmodel}):}} & \tfrac{\partial P}{\partial
t}+\tfrac{u\partial P}{\partial x}+\tfrac{\gamma P\partial u}{\partial
x}=0, &
\begin{array}
[c]{c}%
\text{{\small thermal pressure of cool inertial electron; }}\\
\text{{\small Eqs. (\ref{eq2_9}) and (\ref{eq2_10})}}%
\end{array}
\\
&  & \\
\text{{\small Model 3 (Chapter \ref{twofluidmodel}):}} & \left\{
\begin{array}
[c]{c}%
\tfrac{\partial\tilde{n}}{\partial t}+\tfrac{\partial(\tilde{n}\tilde{u}%
)}{\partial x}=0,\\
\\
\tfrac{\partial\tilde{u}}{\partial t}+\tfrac{\tilde{u}\partial\tilde{u}%
}{\partial x}=-\tfrac{\mu\partial\phi}{\partial x},
\end{array}
\right.  &
\begin{array}
[c]{c}%
\text{{\small ion inertial continuity }}\\
\text{{\small and momentum equations; }}\\
\text{{\small Eqs. (\ref{eq3_12}) and (\ref{eq3_13})}}%
\end{array}
\end{array}
\nonumber
\end{equation}%
\begin{equation}%
\begin{array}
[c]{cc}%
\dfrac{\partial^{2}\phi}{\partial x^{2}}=\left\{
\begin{array}
[c]{cc}%
-(1+\beta) &
\begin{array}
[c]{c}%
\text{{\small inertialess ions}}\\
\text{{\small Chapters \ref{singlecold} and \ref{warmmodel}}}%
\end{array}
\\
-\tilde{n} &
\begin{array}
[c]{c}%
\text{{\small inertial ions}}\\
\text{{\small Chapter \ref{twofluidmodel}}}%
\end{array}
\end{array}
\right.  +\underset{\text{{\small Chapters \ref{singlecold},\ref{warmmodel},
and \ref{twofluidmodel}}}}{\underbrace{\underset{%
\begin{array}
[c]{c}%
\text{{\small cool }}\\
\text{{\small electrons}}%
\end{array}
}{\underbrace{n}}+\underset{\text{{\small hot suprathermal electrons}}%
}{\underbrace{\beta\left(  1-\dfrac{\phi}{(\kappa-\tfrac{3}{2})}\right)
^{-\kappa+1/2}}}},} &
\begin{array}
[c]{c}%
\text{{\small Poisson's equation}}\\
\text{{\small Eqs. (\ref{eq1_12}), (\ref{eq2_29}), and (\ref{eq3_14})}}%
\end{array}
\end{array}
\nonumber
\end{equation}
\end{landscape}

\chapter{An Alternative to Two-Fluid Model: Ion Temperature Effects}\label{hotionseffect}

We may also consider the ion thermal pressure. Due to the thermal pressure of
the ions, the equation of momentum contains an extra term (compare to Eq.
(\ref{eq3_5}))
\begin{equation}
\frac{\partial u_{i}}{\partial t}+u_{i}\frac{\partial u_{i}}{\partial
x}=-\frac{Ze}{m_{i}}\frac{\partial\phi}{\partial x}-\frac{1}{m_{i}n_{i}}
\frac{\partial P_{i}}{\partial x}, \label{eq4_1}
\end{equation}
The pressure of the ions is given by
\begin{equation}
\frac{\partial P_{i}}{\partial t}+u_{i}\frac{\partial P_{i}}{\partial
x}+\gamma P_{i}\frac{\partial u_{i}}{\partial x}=0, \label{eq4_2}
\end{equation}
where $P_{i}$ is the thermal pressure of the ions, in the one-dimensional
model $\gamma=3$. We define the temperature ratio of the ions to the hot
electrons as $\tilde{\sigma}=T_{i}/T_{h}$.

The normalized forms of Eqs. (\ref{eq4_1})--(\ref{eq4_2}) are written as
($Z=1$ and $\gamma=3$):
\begin{equation}
\frac{\partial\tilde{u}}{\partial t}+\tilde{u}\frac{\partial\tilde{u}
}{\partial x}=-\mu\frac{\partial\phi}{\partial x}-\frac{\mu\tilde{\sigma}
}{\tilde{n}}\frac{\partial\tilde{P}}{\partial x} \label{eq4_3}
\end{equation}
\begin{equation}
\frac{\partial\tilde{P}}{\partial t}+\tilde{u}\frac{\partial\tilde{P}
}{\partial x}+3\tilde{P}\frac{\partial\tilde{u}}{\partial x}=0, \label{eq4_4}
\end{equation}
The density $n_{i}$ are normalized with the unperturbed cool density
($n_{c,0}$), the velocity $u_{i}$ with the hot electron thermal velocity
($c_{h,s}=\left(  k_{B}T_{h}/m_{e}\right)  ^{1/2}$), time with the inverse
cool electron plasma frequency, $\omega_{pc}^{-1}$, where $\omega
_{pc}=(n_{c,0}e^{2}/\varepsilon_{0}m_{e})^{1/2}$, length with the
characteristic length scale, $\lambda_{0}=(\varepsilon_{0}k_{B}T_{h}
/n_{c,0}e^{2})^{1/2}$, the wave potential $\phi$ with $k_{B}T_{h}/e$, and the
pressure $P_{i}$ with $n_{c,0}k_{B}T_{i}$.

Integrating Eqs. (\ref{eq4_3}) and (\ref{eq4_4}) yield
\begin{equation}
\begin{array}
[c]{cc}
\tilde{u}=M{-}\sqrt{{M}^{2}{-2\mu\phi-3\mu\tilde{\sigma}\left[  \tilde{n}
^{2}-(1+\beta)^{2}\right]  }}, & \text{ \ }\tilde{P}=\tilde{n}^{3}.
\end{array}
\label{eq4_5}
\end{equation}
Combining Eqs. (\ref{eq3_40}a)--(\ref{eq4_5}), we get
\begin{equation}
{\tilde{n}=}\frac{1}{2}\sqrt{\frac{\left(  {M+{(1+\beta)}}\sqrt{{{3\mu
\tilde{\sigma}}}}\right)  ^{2}{-2\mu\phi}}{3{{\mu\tilde{\sigma}}}}}\pm\frac
{1}{2}\sqrt{\frac{\left(  {M-{(1+\beta)}}\sqrt{{{3\mu\tilde{\sigma}}}}\right)
^{2}{-2\mu\phi}}{3{{\mu\tilde{\sigma}}}}} \label{eq4_7}
\end{equation}

Therefore, the Sagdeev pseudopotential (\ref{eq3_49}) is rewritten as
\begin{align}
& \Psi_{4}(\phi,M,\beta,\kappa,{\sigma},\tilde{\sigma},\mu)    =-\beta\left(
\left(  1+\frac{\phi}{-\kappa+\tfrac{3}{2}}\right)  ^{-\kappa+3/2}-1\right)
+M^{2}+\sigma\nonumber\\
&  -\frac{1}{6\sqrt{3\sigma}}\left[  {2\phi+}\left(  {M+\sqrt{3{\sigma}}
}\right)  ^{2}\right]  ^{3/2}\mp\frac{1}{6\sqrt{3\sigma}}{\left[
{2\phi+\left(  {M-\sqrt{3{\sigma}}}\right)  ^{2}}\right]  }^{3/2}\nonumber\\
&  +\frac{(1+\beta)}{{\mu}}\left[  M^{2}+{(1+\beta)}^{2}\mu\sigma\right]
-\frac{1}{6{\mu}\sqrt{3{{\mu\tilde{\sigma}}}}}\left[  \left(  {M+{(1+\beta)}
}\sqrt{{{3\mu\tilde{\sigma}}}}\right)  ^{2}{-2\mu\phi}\right]  ^{3/2}
\nonumber\\
&  \mp\frac{1}{6{\mu}\sqrt{3{{\mu\tilde{\sigma}}}}}\left[  \left(
{M-{(1+\beta)}}\sqrt{{{3\mu\tilde{\sigma}}}}\right)  ^{2}{-2\mu\phi}\right]
^{3/2} \label{eq4_8}
\end{align}
In the limit $\tilde{\sigma}\rightarrow0$, we get Eq. (\ref{eq3_49}).

For the existence of acoustic-solitary waves moving at velocity $M$, we
require $\Psi_{4}^{\prime}(\phi,M,\beta,\kappa,{\sigma},\tilde{\sigma}
,\mu)|_{\phi=0}=0$ and $F_{1}(M,\beta,\kappa,{\sigma},\tilde{\sigma}
,\mu)\equiv-\Psi_{4}^{\prime\prime}(\phi,M,\beta,\kappa,{\sigma},\tilde
{\sigma},\mu) $ $|_{\phi=0}>0$. Here, the function $F_{1}(M)$ reads as
\begin{equation}
F_{1}(M,\beta,\kappa,{\sigma},\tilde{\sigma},\mu)=\frac{\beta(\kappa-\frac
{1}{2})}{\kappa-\tfrac{3}{2}}-\frac{1}{M^{2}-3\sigma}-\frac{{(1+\beta)}\mu
}{{M}^{2}{-{(1+\beta)}}^{2}{{3\mu\tilde{\sigma}}}}, \label{eq4_9}
\end{equation}
where ${{(1+\beta)}}\sqrt{{{3\mu\tilde{\sigma}}}}$ is the normalized ion
thermal velocity.

\begin{figure}
\begin{center}
\includegraphics[width=5.0in]{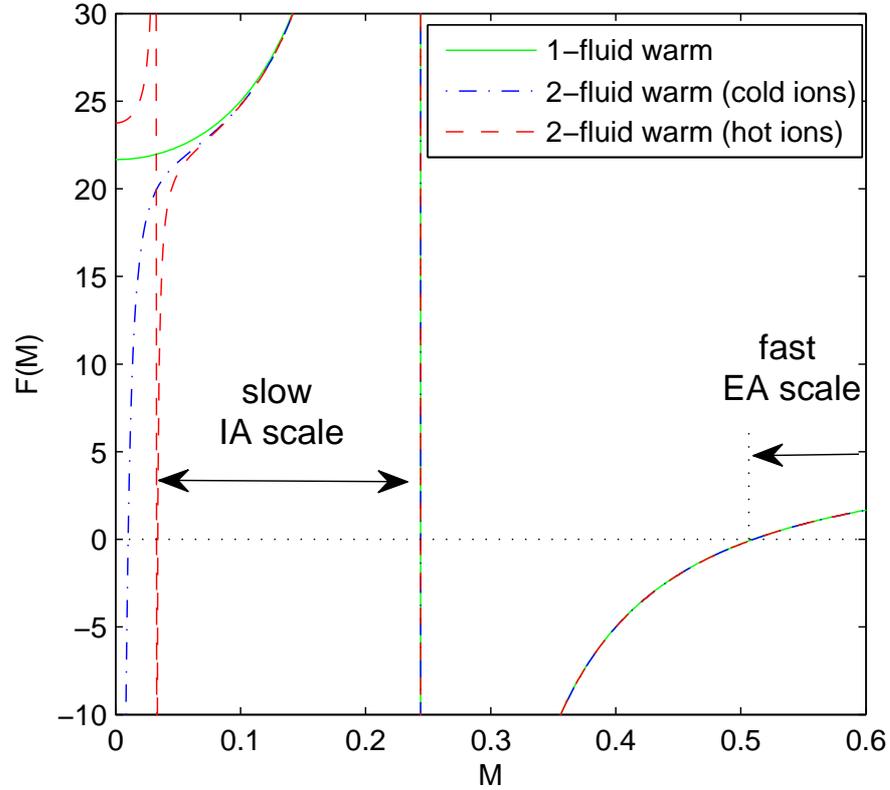}
\caption[The existence domains for stationary solitary structures.]{The existence domains for stationary solitary structures. The
quantities $F_{1}$ for 1-fuild warm model (solid), 2-fluid warm model for
$T_{i}=0$ (dot-dashed), and 2-fluid cold model for $T_{i}\neq0$ (dashed
curve), are defined in (\ref{eq2_44}), (\ref{eq3_56}), and (\ref{eq4_9})
respectively. Here, $\kappa=3$, $\beta=3$, $\sigma=0.02$, $\tilde{\sigma
}=0.04$, $Z=1$, and $\mu=1/1836$. }
\label{fig4_3}
\end{center}
\end{figure}

We see that Eq. (\ref{eq4_9}) contains an extra term corresponding to the ion
thermal pressure (compare to Eq. (\ref{eq3_56})). The ion thermal velocity
classifies the Mach number under two regions, namely \textquotedblleft cool
ion\textquotedblright\ (${M>{(1+\beta)}}\sqrt{{{3\mu\tilde{\sigma}}}}$) and
\textquotedblleft hot ion\textquotedblright\ (${M<{(1+\beta)}}\sqrt
{{{3\mu\tilde{\sigma}}}}$), in the sense that the thermal velocity is
smaller/larger than $M$, respectively. As illustrated in Fig. \ref{fig4_3},
including the hot ion component ($T_{i}\neq0$) divides the propagation speed
into three ranges. Nonetheless, there are two existence ranges for solitary
waves as Chapter \ref{twofluidmodel}. The existence range for positive
acoustic-solitary waves has been effectively changed.

\chapter{Solving Biquadratic Equation}\label{BiquadraticEquation}

The quartic equation takes the form as
\begin{equation}
Q(x)=a_{4}x^{4}+a_{3}x^{3}+a_{2}x^{2}+a_{1}x+a_{0}.
\end{equation}
If $a_{3}=a_{1}=0$, then we get the biquadratic equation
\begin{equation}
Q(x)=a_{4}x^{4}+a_{2}x^{2}+a_{0}. \label{biquadratic}
\end{equation}
Let assume $x=\sqrt{x_{1}}\pm\sqrt{x_{2}}$, we get $x^{2}=x_{1}+x_{2}\pm
2\sqrt{x_{1}x_{2}}$, and
\begin{equation}
\begin{array}
[c]{cc}
x_{1}+x_{2}=-\dfrac{a_{2}}{2a_{4}}, & \text{ \ }x_{1}-x_{2}=\sqrt{\dfrac
{a_{0}}{a_{4}}}.
\end{array}
\end{equation}
We then find the following solution to the biquadratic equation
(\ref{biquadratic}):
\begin{equation}
x=\sqrt{-\dfrac{a_{2}}{4a_{4}}+\dfrac{1}{2}\sqrt{\dfrac{a_{0}}{a_{4}}}}
\pm\sqrt{-\dfrac{a_{2}}{4a_{4}}-\dfrac{1}{2}\sqrt{\dfrac{a_{0}}{a_{4}}}}.
\end{equation}

\cleardoublepage







\end{document}